\documentclass[prl,twocolumn,epsfig]{revtex4}
\usepackage{graphicx}
\usepackage{dcolumn}
\usepackage{bm}

\begin{document}
\title {Nanoscale Electrostatic Control of Oxide Interfaces}
\author{Srijit~Goswami}
\email{s.goswami@tudelft.nl}
\affiliation{Kavli Institute of Nanoscience, Delft University of Technology, P.O. Box 5046, 2600 GA Delft, The Netherlands}
\author{Emre~Mulazimoglu}
\affiliation{Kavli Institute of Nanoscience, Delft University of Technology, P.O. Box 5046, 2600 GA Delft, The Netherlands}
\author{Lieven~M.~K.~Vandersypen}
\affiliation{Kavli Institute of Nanoscience, Delft University of Technology, P.O. Box 5046, 2600 GA Delft, The Netherlands}
\author{Andrea~D.~Caviglia}
\affiliation{Kavli Institute of Nanoscience, Delft University of Technology, P.O. Box 5046, 2600 GA Delft, The Netherlands}

\begin{abstract}

We develop a robust and versatile platform to define nanostructures at oxide interfaces via patterned top gates. Using LaAlO$_3$/SrTiO$_3$ as a model system, we demonstrate controllable electrostatic confinement of electrons to nanoscale regions in the conducting interface. The excellent gate response, ultra-low leakage currents, and long term stability of these gates allow us to perform a variety of studies in different device geometries from room temperature down to 50~mK. Using a split-gate device we demonstrate the formation of a narrow conducting channel whose width can be controllably reduced via the application of appropriate gate voltages. We also show that a single narrow gate can be used to induce locally a superconducting to insulating transition. Furthermore, in the superconducting regime we see indications of a gate-voltage controlled Josephson effect.

\end{abstract}

\maketitle

Despite decades of intense study, transition metal oxides continue to reveal fascinating and unexpected physical properties that arise from their highly correlated electrons~\cite{Dagotto_Science_Rev}. Propelled by recent developments in oxides thin film technology it has now become possible to create high quality interfaces between such complex oxides, which reveal a new class of emergent phenomena often non-existent in the constituent materials~\cite{Zubko_Triscone_ARCMP, Hwang_NatMat_Rev}. In particular, there has been a growing interest in interfaces that host a conducting two dimensional electron system (2DES)~\cite{Ohtomo_Nature, Chen_Al2O3_NatComm}. This 2DES has been shown to support high mobility electrons~\cite{Caviglia_SdH_PRL, Huijben_Adv_Fun_Mat, Chen_Al2O3_NatComm}, magnetism~\cite{Brinkman_Nat_Mat} and superconductivity~\cite{Reyren_Science}. In addition to this inherently rich phase space, in-situ electrostatic gating can be used not only to alter the carrier density~\cite{Thiel_Science}, but it can significantly change the spin-orbit coupling (SOC)~\cite{Caviglia_SO_PRL, Shalom_SO_PRL} and even drive transitions from a superconducting to an insulating state~\cite{Caviglia_Nature}.

Bulk transport studies of oxide interfaces have played an important role toward building a better understanding of these new material systems. However, it is becoming increasingly clear that in order to fully grasp the details of the complex coexisting phases at the interface, one must probe the system at much smaller length scales. Recent scanning probe experiments have indeed clearly demonstrated that the electronic properties of the interface can change dramatically over microscopic length scales~\cite{Bert_Moler_Nat_Phys, Kalisky_Moler_Nat_Mat, Honig_Ilani_Nat_Mat}. In this context, nanoscale electronic devices could provide direct information on how such strong local variations in physical properties affect mesoscopic charge transport. Perhaps even more exciting is the possibility of discovering and manipulating new electronic states that are predicted to arise from the interplay between confinement, superconductivity and SOC~\cite{Fidkowski_Majo_PRB}. Furthermore, the ability to locally drive phase transitions at the interface could potentially yield technologically relevant oxide-based nano-electronic devices with novel functionality~\cite{Mannhart_Science_Rev}.

Existing methods for confinement at the interface involve some form of nanoscale patterning, which renders selected portions of the interface insulating, while others remain conducting. These include the use of pre-patterned masks~\cite{Stornaiuolo_confinement_APL}, physical etching of the interface~\cite{Ionbeam_APL}, and AFM-based lithography~\cite{Cen_NatMat,Cen_Science, Cheng_Levy_SET_NatNano}. Such techniques have shown promising results, and have been used to realize transistor-like nanoscale devices controlled via the field effect from a global back gate or local side gates. However they do not provide an obvious way to create more intricate device structures that require in-situ tunability of the potential landscape. In addition, they suffer from issues such as ion-beam induced damage or long term stability, which could have a direct impact on device performance.

These hurdles can be overcome by the use of local top gates, which can be conveniently integrated with several oxide interfaces where the top oxide layer itself acts as a high quality gate dielectric. In this device architecture the potential profile in the 2DES can be precisely controlled using appropriate gate voltages, thus making it an extremely flexible and robust platform to build tailor-made nanostructures. Such electrostatic confinement is routinely employed to create low dimensional systems in traditional semiconductor based 2DESs. It is therefore somewhat surprising that a similar strategy has not been adopted to investigate confinement at oxide interfaces thus far. Large area top gated devices have indeed been fabricated using a variety of techniques such as sputtering~\cite{Hosada_APL}, evaporation~\cite{Eerkes_APL,Jany_AdvMat} and in-situ deposition~\cite{Forg_APL, Richter_Nature_Gap}. However scaling these structures down has remained a challenge.

Here, we define nanoscale electronic devices in LaAlO$_3$/SrTiO$_3$ (LAO/STO) using patterned top gates that efficiently modify the potential landscape at the metallic interface. We demonstrate that individual narrow gates (down to 200 nm) can completely pinch off the conducting channel and display large on/off ratios with negligible leakage currents. Using two such gates in a split-gate geometry, we can further tune the flow of charge carriers by restricting them to a narrow conducting channel with a width that can be controlled in-situ via the gate voltages. At milliKelvin temperatures, when the interface is superconducting we use a single narrow top gate to drive locally a superconducting  to insulating transition at the LAO/STO interface. In the superconducting state, we see evidence for a superconducting weak link with a gate-dependent critical current.

Device fabrication involves pulsed laser deposition for the oxide growth in combination with multiple aligned lithography steps (see SI for device specific details). Single crystal STO~(001) substrates first undergo photolithography or electron beam lithography (EBL), followed by the deposition of 45~nm of amorphous LAO (a-LAO). The a-LAO is deposited at room temperature with an O$_2$ pressure of $6\times 10^{-5}$~mbar and laser fluency of 1~J/cm$^2$ (repetition rate: 5~Hz). Subsequent lift-off in warm acetone (50~$^\circ$C) creates an a-LAO mask for the crystalline LAO (c-LAO) deposition. We deposit 12 unit cells of c-LAO at 770~$^\circ$C with an O$_2$ pressure of $6\times 10^{-5}$~mbar and laser fluency of 1~J/cm$^2$ (repetition rate: 1~Hz). The film growth is monitored in-situ using reflection high-energy electron diffraction (RHEED) confirming layer by layer growth. Finally, a one hour long post-growth anneal is performed at 300~mbar O$_2$ pressure and 600~$^\circ$C, followed by a cooldown to room temperature in the same atmosphere. Under these conditions the 2DES formed at the LAO/STO interface shows sheet densities of about 3$\times$10$^{13}$~cm$^{-2}$ and field effect mobilities up to 3500~cm$^2$/Vs at 4.2~K in bulk samples. The patterned top gates are finally defined by an aligned EBL step, followed by electron beam evaporation of 100~nm Au directly on the c-LAO surface.

\begin{figure}[!t]
\includegraphics[width=1\linewidth]{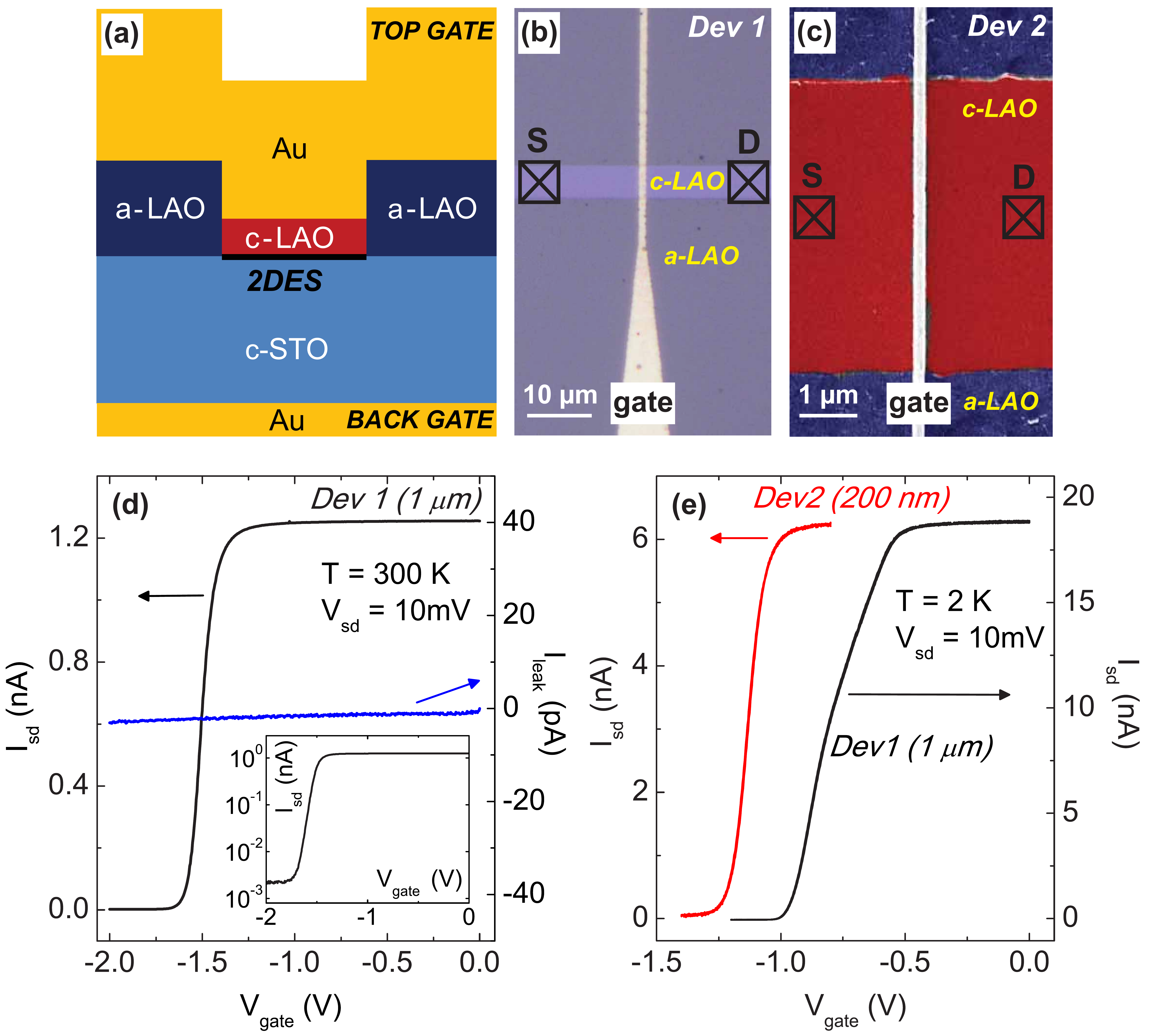}
\caption{(a) Cross-sectional schematic of a typical top gated LAO/STO device. The 2DES (indicated in black) is formed at the interface between crystalline LAO (c-LAO) and STO (c-STO). The interface between amorphous LAO (a-LAO) and STO remains insulating. (b) Optical image of Dev1 with a 1~$\mu$m wide gate running over the c-LAO. (c) False color scanning electron micrograph of Dev2, which consists of a 200~nm wide top gate. Red/blue areas correspond to conducting/insulating regions. See SI for full device drawings with the relevant channel dimensions. (d) Gate characteristics for Dev1 at 300~K. Black curve (left axis) shows the variation of source drain current ($I_{sd}$) with top gate voltage ($V_{gate}$). Inset: the same curve on a logarithmic scale. Blue curve (right axis) shows the leakage current ($I_{leak}$) \emph{vs.} $V_{gate}$. (e) Gate response of Dev1 and Dev2 at 2~K. Each curve consists of 10 consecutive down sweeps from the on to off state.}
\label{fig1}
\end{figure}

Figure~\ref{fig1}a shows a cross-sectional schematic of a device with a single top gate. An optical image of such a device is shown in Figure~\ref{fig1}b, where a narrow (1~$\mu$m wide) gate runs across a mesoscopic conducting channel defined at the LAO/STO interface. Figure~\ref{fig1}d shows the gate characteristics of this device (Dev1) at 300~K (see SI for data from similar devices). A constant dc voltage bias of 10~mV is applied across S and D ($V_{sd}$) and the current ($I_{sd}$) is measured as a function of the top gate voltage (V$_{gate}$). The black curve (left axis) shows a typical field effect behavior with complete depletion under the gated region resulting in an on/off ratio of nearly 1000 (shown in the inset). The leakage current (blue trace, right axis) remains below 5~pA in this gate voltage range, thus allowing for reliable measurements of very low currents through the device. The gate width can be reduced even further (in this case to 200~nm), as seen in the (false color) scanning electron microscope (SEM) image of Dev2 (Figure~\ref{fig1}c). In Figure~\ref{fig1}e we compare the gate response of Dev1 and Dev2 at $T=2$~K. Both devices (Dev1-black trace, right axis; Dev2-red trace, left axis) show comparable threshold voltages. This is consistent with the fact that they were both fabricated on 12 unit cell LAO deposited under the same growth conditions. We note that each curve consist of 10 consecutive sweeps between the on and off states. It is clear that the device completely recovers from the insulating state, and is extremely stable over multiple on/off cycles. This is in contrast with recent measurements on bulk top gated LAO/STO devices where (at low temperatures) going above a critical resistance rendered the interface completely insulating, and conduction could only be revived by thermal cycling~\cite{Eerkes_APL}. We point out that the two-terminal resistance values are significantly higher than those expected from the bulk mobility and density values mentioned earlier. This discrepancy most likely arises from a contact resistance and/or local regions in the long narrow channel with a significantly higher resistivity.

\begin{figure*}[!t]
\includegraphics[width=1\linewidth]{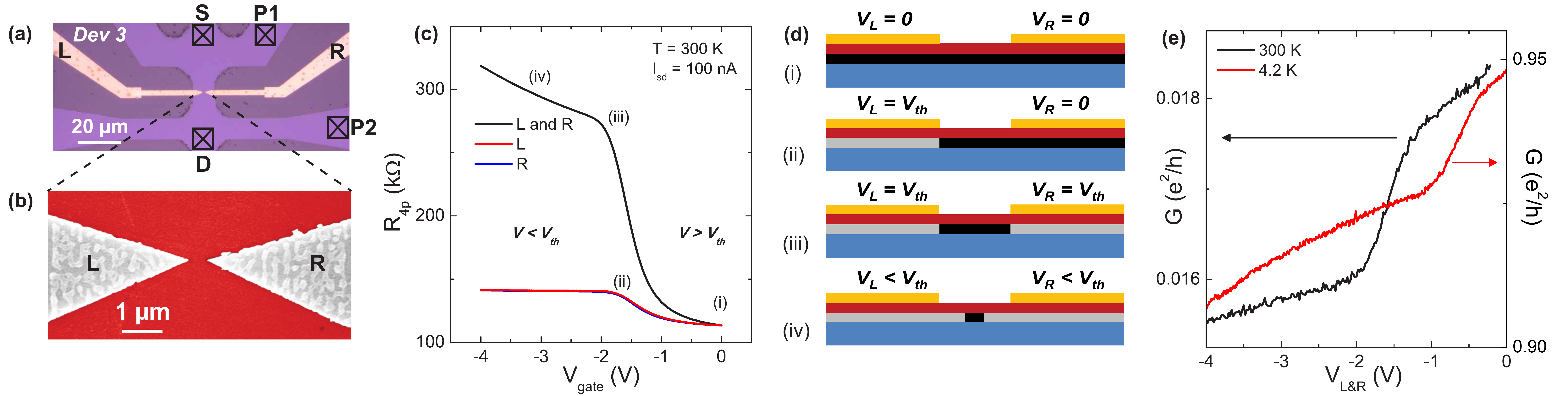}
\caption{(a) Optical and (b) SEM images of a split gate (SG) device which comprises of a left (L) and right (R) gate. (c) Four-probe resistance ($R_{4p}$) measurements (at 300~K) comparing the gate action of the individual gates (red and blue curves) with that of both gates together (black curve). Note that the red and blue curves show an excellent overlap, which makes it hard to distinguish between the two in the plot. The labels [(i)-(iv)] mark different transport regimes in the operation of the SG device. The corresponding density variations at the LAO/STO interface are shown schematically in (d), where black/gray represent conducting/insulating regions in the 2DES. (e) Two-probe voltage biased measurements: Conductance, $G$ \emph{vs.} $V_{L\&R}$ at 300~K (black curve) and 4.2~K (red curve). See SI for device drawings with the relevant channel dimensions.}
\label{fig2}
\end{figure*}

Having established a reliable gate response from our local top gates we use \emph{two} such gates to realize a split gate (SG) geometry, using which charge carriers can be confined to narrow conducting channels. SGs have been successfully used to fabricate quantum point contacts (QPCs) in semiconductor-based 2DESs~\cite{QPC_vanWees_PRL, QPC_Wharam}, and have provided insights into several aspects of mesoscopic physics (reviewed in~\cite{Beenakker_SSP_Rev}) ranging from the quantum Hall effect to the Aharonov-Bohm effect. They are also extremely sensitive charge detectors~\cite{QPC_ChSensor_Field} and spin filters~\cite{QPC_SpFilter_Folk}, and serve as essential components for the creation of lower dimensional systems such as quantum dots~\cite{Kouwenhoven_DotReview}. At the LAO/STO interface, such one-dimensional (1D) confinement could possibly give rise to exotic electronic states that emerge from the interplay between 1D superconductivity and SOC~\cite{Fidkowski_Majo_PRB}. Figure~\ref{fig2}a shows an optical micrograph of a SG device on LAO/STO. It consists of a left (L) and right (R) gate, both of which start off 2.5~$\mu$m wide and taper down to a narrow point. Figure~\ref{fig2}b shows an SEM image of the active device region. We have studied devices with tip separations of approximately 400~nm (Dev3) and 250~nm (Dev4), and both show qualitatively similar features. Here we focus on Dev3 (results from Dev4 can be found in the SI)

We begin by studying the room temperature four probe resistance ($R_{4p}$) across the SG, as a function of the individual (L, R) gates. A 100~nA dc current is applied between S and D ($I_{sd}$), and resistance is measured between contacts P1 and P2 (see Figure~\ref{fig2}a). For these measurements no back gate voltage was applied. When the SG is held at zero ($V_{L}=V_{R}=0$), the carrier density in the entire channel is uniform. This is shown schematically in Figure~\ref{fig2}d (top panel), where the uniform black area represents a wide channel with no density variations. The corresponding point in the $R_{4p}$ vs $V_{gate}$ plot (Figure~\ref{fig2}c) is marked by the label (i). As one of the gates (say L) is made more negative, the carriers below this gate are depleted, resulting in a sharp increase in $R_{4p}$ (red curve). However, at a critical threshold voltage ($V_{th}$) the region below R is completely depleted, indicated by (ii) in Figure~\ref{fig2}c,d (gray areas represent depleted regions in the 2DES). Throughout, we define $V_{th}$ as the gate voltage where the magnitude of this slope is maximum. Beyond $V_{th}$ the gate action is weaker, since depletion must now occur sideways, thereby resulting in a lower slope in the $R_{4p}$ \emph{vs.} $V_{gate}$ curve. Sweeping only L (instead of R) should be electrostatically equivalent to the situation described above. This is reflected directly in transport by the excellent overlap between the red and blue curves. If \emph{both} R, L are swept together the response is much stronger (black curve). When $V_{L}=V_{R}=V_{th}$ [label (iii)] a narrow constriction is formed in the 2DES, whose width is determined by the geometry of the split gates and the electrostatics of the system. Finally, going to even more negative voltages with L and R together [label (iv)] squeezes this channel further. Thus, through an appropriate device design and suitable gate voltages, it is clearly possible to electrostatically define nanoscale constrictions at an oxide interface, even at room temperature.

Next, we study this SG device (Dev3) at cryogenic temperatures in a two probe configuration. Figure~\ref{fig2}e shows the variation of the conductance ($G$) with $V_{L\& R}$ (i.e., both L and R swept together) at 300~K (black curve, left axis) and 4.2~K (red curve, right axis). Two things are immediately apparent from these plots. First, the conductivity of the system increases significantly as the temperature is lowered, which is expected for such metallic samples. Second, $V_{th}$ shifts to less negative voltages upon cooling down. It has been suggested that the sheet density can reduce with temperature as a result of carrier freeze-out~\cite{Huijben_Adv_Fun_Mat}. Such a reduction in the sheet density with temperature could qualitatively explain the observed shift in $V_{th}$, since it now becomes much easier for the SG to deplete the carriers. However, at these temperatures an even more striking effect appears in the gate response of the SGs, which only becomes apparent through detailed phase space maps of the conductance \emph{vs.} the individual gates.

\begin{figure}[!t]
\includegraphics[width=1\linewidth]{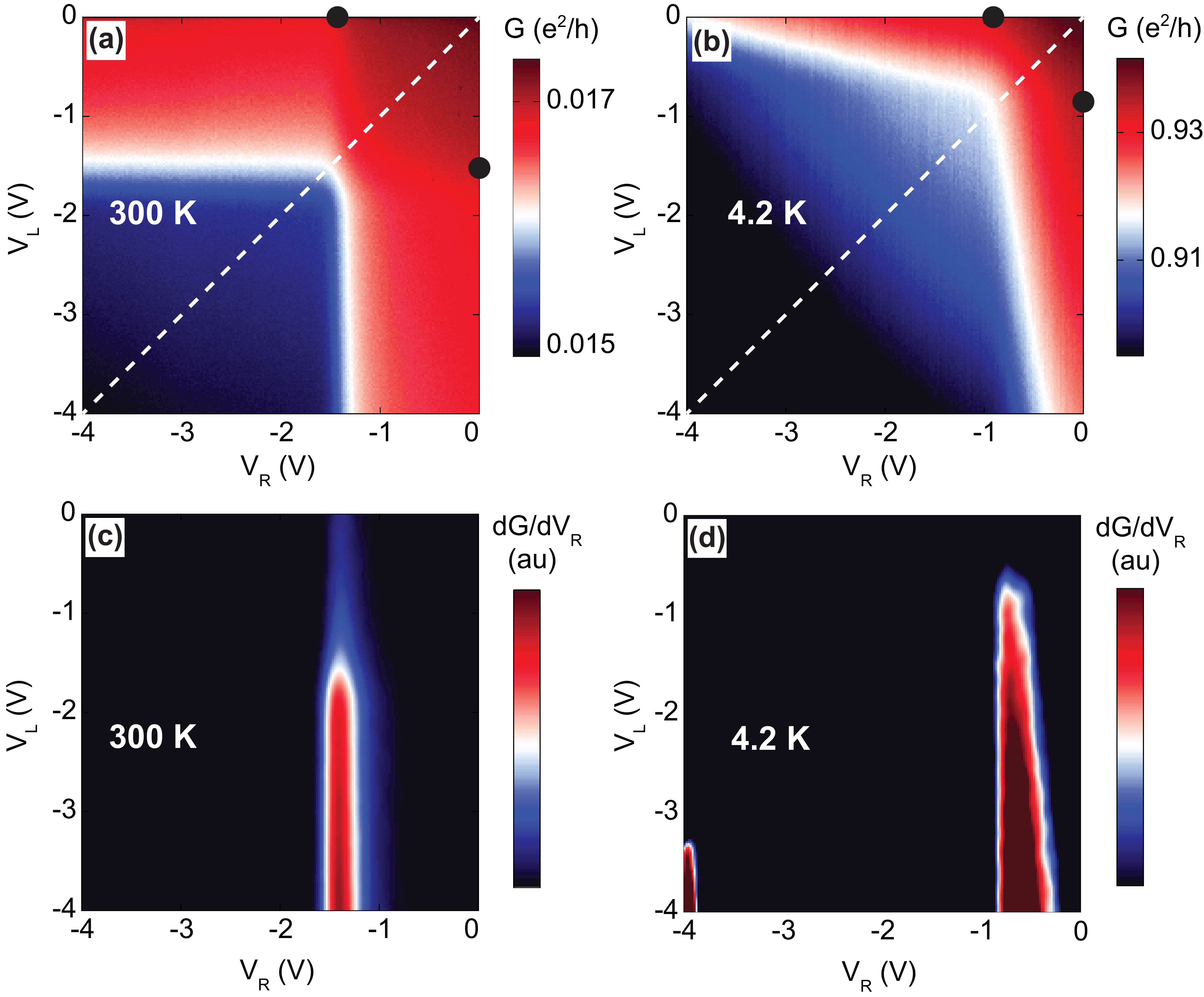}
\caption{2D maps of conductance ($G$) as a function of left/right (L/R) split gate voltages ($V_L$/$V_R$) at (a) $T=300$~K and (b) $T=4.2$~K. The gate voltage step (for both $V_L$ and $V_R$) is 20~mV. Black circles mark the positions of the threshold voltage for either gate, with the other held at zero. Symmetry about the white dashed line indicates comparable gate action from both L and R. (c)\&(d) show plots of the corresponding numerical derivatives taken along the $V_R$ axis ($dG/dV_R$). The maximum in $dG/dV_R$ indicates the threshold voltage for gate R. It is independent of $V_L$ at 300~K, but shows a distinct shift at 4.2~K, indicating cross-talk between L and R at low temperatures.}
\label{fig3}
\end{figure}

Figure~\ref{fig3}a,b show two such maps of $G$, at 300~K and 4.2~K respectively. It is worth pointing out that these maps are typically acquired over several hours, during which the device does not show any switches or obvious drifts, confirming the stability and robustness of these gates. Furthermore, both gates (L and R) have a nearly identical influence on the 2DES at the interface. This is evident from the high degree of symmetry across the diagonal (white dashed line). Such a symmetric response of the two gates can be expected if the 2DES has a homogeneous density. However, inhomogeneities in the gated regions could give rise to a reduction in this symmetry (as seen for Dev4 in the SI). The entire phase space can be divided into two distinct regions. The blue portion (lower $G$) corresponds to a situation when both L and R have been driven beyond their respective threshold voltages. The red area (higher $G$), thus reflects  the complementary scenario, where either one (or none) of the gates have crossed $V_{th}$ (black circles indicate the position of $V_{th}$ for each of the gates with the other held at zero). The narrow white band therefore separates these two electrostatically distinct regimes and provides information about the effective region of influence of the individual gates, and the extent of cross-talk between them. These effects can be examined in a more consistent manner by taking a derivative along one of the gate axes.

Figure~\ref{fig3}c,d show the corresponding numerical derivatives taken along the $V_{R}$ axis ($dG/dV_{R}$). As mentioned earlier, the maximum in $dG/dV_{R}$ occurs at $V_{th}$ associated with gate~R. At 300~K, as $V_{L}$ is made more negative the position of the threshold does not change, indicating that there is practically no cross-talk between the gates. This is perhaps not so surprising considering the fact that the gates are very close ($\sim$5~nm) to the interface, and therefore most of the electric field lines go directly downward, creating a sharp potential profile with minimal spreading of the the electric field. However, at 4.2~K the situation is rather different. $V_L$ obviously has a distinct effect on the threshold voltage for R, shifting it to less negative voltages, as $V_L$ becomes more negative. This suggests that $V_L$ has a significant influence on the region below R, which is in contrast with the observations at 300~K. Such a modification of the electrostatics could possibly be related to the fact that STO is an incipient ferroelectric, thereby exhibiting a strong increase in its permittivity at low temperatures~\cite{Edep_dielectric}. A combination of this large permittivity ($\sim 10^4$ at $T=4$~K) along with imperfect screening from the 2DES, could provide a viable mechanism for the observed cross-talk between the gates. As the carrier density below L is reduced, the extent of screening from the 2DES below L reduces. This, in turn, allows electric field lines to go \emph{through} the STO, resulting in a significant field effect in the region below R. Though a likely explanation, the feasibility of such a scenario ultimately needs to be tested using electrostatic simulations with appropriate parameters for the screening lengths at the LAO/STO interface.
\begin{figure}[!b]
\includegraphics[width=1\linewidth]{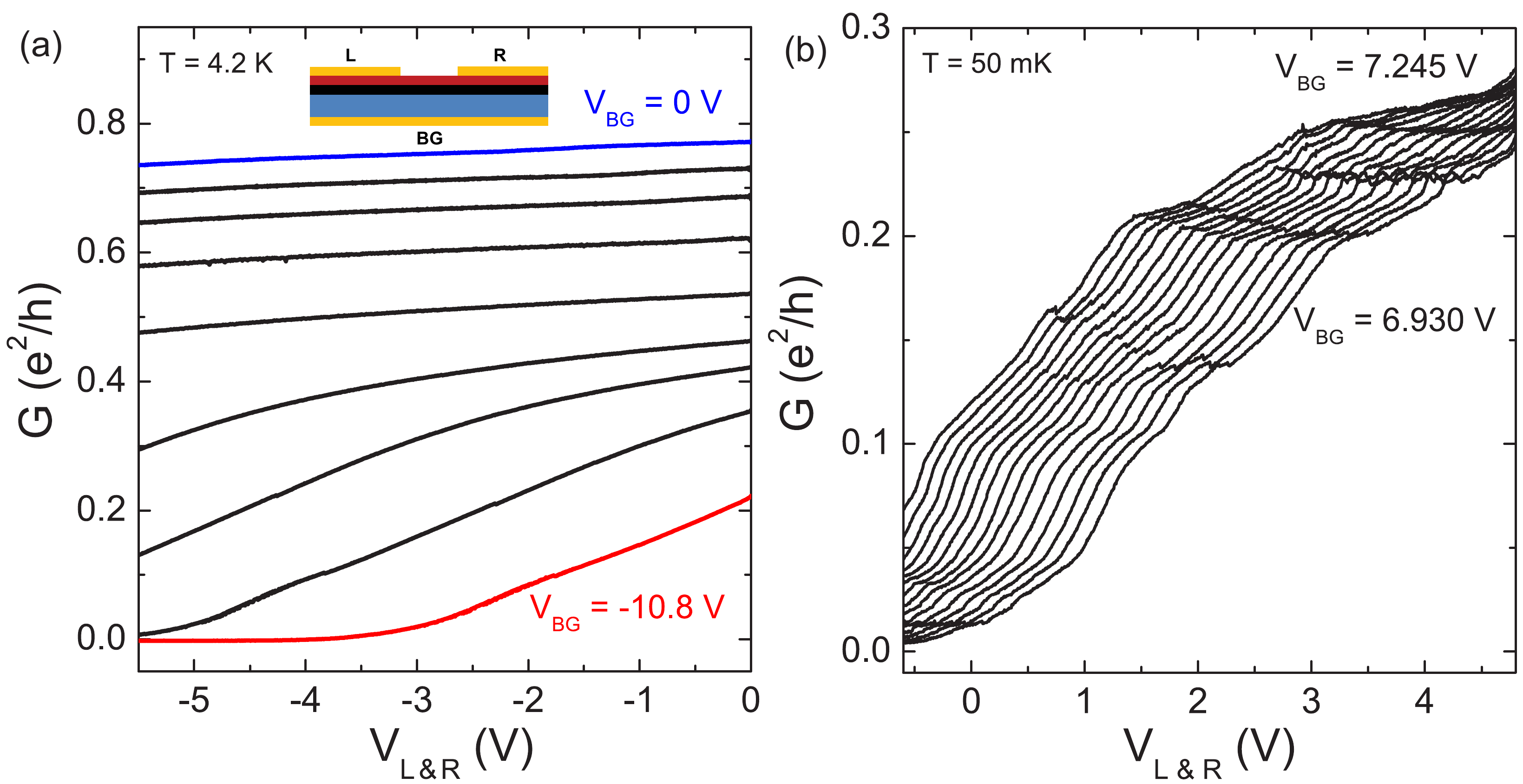}
\caption{(a) The conductance ($G$) of the nano-constriction formed by the split gates can also be tuned using the back gate (BG). While $G$ does not change appreciably with $V_{L\&R}$ when $V_{BG}=0$~V (blue curve, a similar trace is also shown in Figure~2e), for $V_{BG}=-10.8$~V it is possible to completely deplete the channel using split gates (red curve). Intermediate (black) curves correspond to $V_{BG}=-1.8$~V, -3.6~V, -5.85~V, -8.1~V, -9.45~V, -9.9~V, and -10.35~V. (b) Modulation of $G$ with $V_{L\&R}$ at milliKelvin temperatures due to mesoscopic effects in the conducting channel. The modulations evolve continuously as a function of BG.}
\label{fig4}
\end{figure}

\begin{figure*}[!t]
\includegraphics[width=1\linewidth]{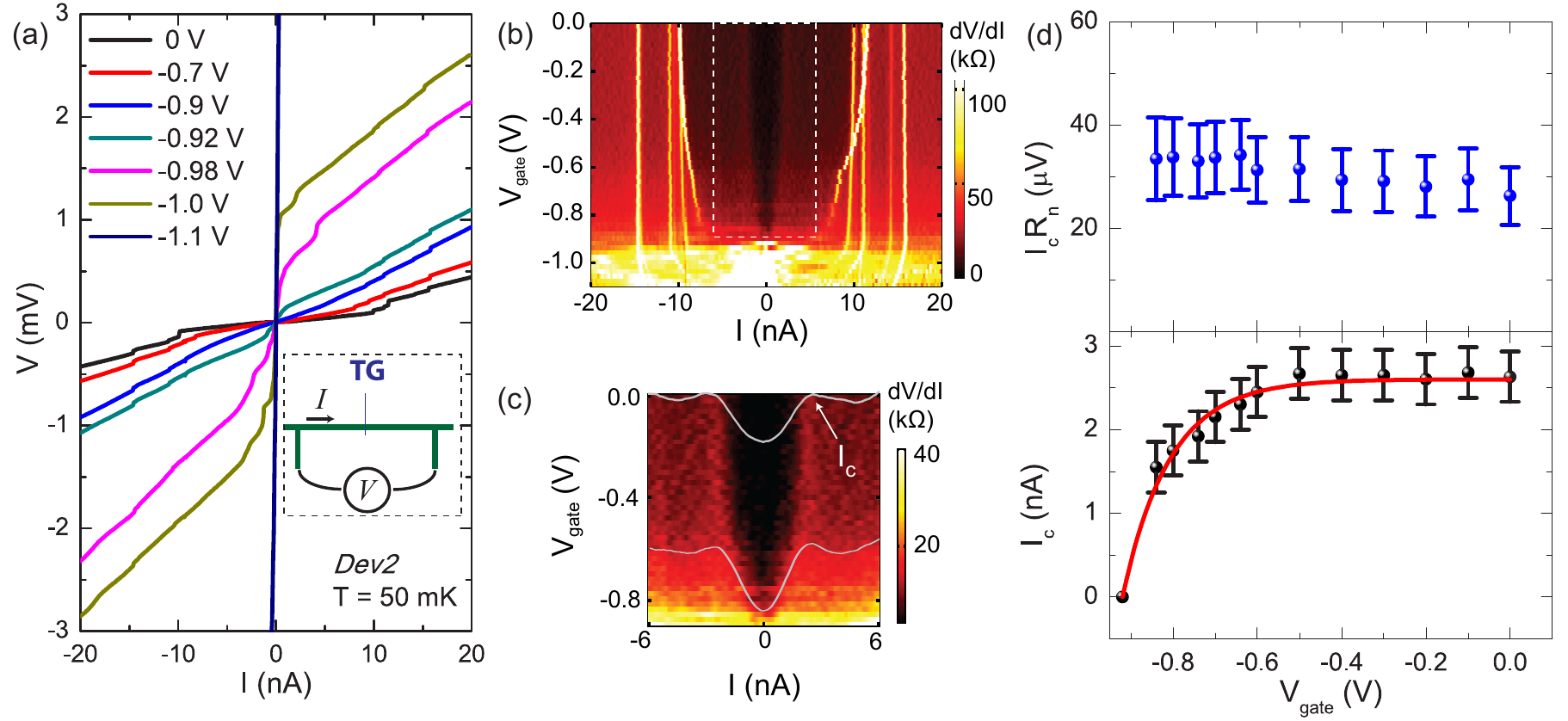}
\caption{(a) A local superconducting to insulating transition induced by the top gate. Inset: schematic (to-scale) of the device and measurement configuration. (b) 2D map of differential resistance ($dV/dI$) as a function of $I$ and $V_{gate}$. (c) A closer view of the region enclosed by the dashed lines in (b), with line traces at $V_{gate}=0$~V and $V_{gate}=-0.6$~V  (d) Lower and upper panels show the critical current ($I_c$) and $I_cR_n$ product respectively of a gate-tunable weak link. Red line is a guide to the eye.}
\label{fig5}
\end{figure*}
In addition to local top gates, the global back gate can also modify the electrostatics at the interface, thus providing additional control over transport through electrostatically defined nanostructures. Though our SG devices show clear evidence of confinement, the relatively large separation between the gates (as compared to the distance of the gates from the 2DES) makes it difficult to pinch off the channel using just the top gates. However, by reducing the sheet carrier density via moderate negative voltages on the back gate (BG), we could indeed deplete the constriction completely, as seen in Figure~\ref{fig4}a. In this geometry the entire 0.5~mm thick STO substrate is used as the gate dielectric. At $V_{BG}=0~V$ the channel remains fairly open in this gate voltage range (blue trace), but shows a clear pinch-off for $V_{BG}=-10.8~V$ (red trace). We note that in all the devices studied here, we observed a strong hysteresis in the BG action at low temperatures. As the BG is taken to positive values, $G$ typically saturates for $V_{BG}>20$~V. Subsequently, going back to $V_{BG}=0$~V renders the sample highly insulating. However, the conductance can be completely recovered by taking the top gate to positive voltages, which suggests that the effect arises primarily from the region below the local gates. Such an effect was not observed in larger Hall bars (channel width~$\sim$~500~$\mu$m), fabricated in a similar fashion, but without top gates.

Figure~\ref{fig4}b shows the variation of $G$ with $V_{L\&R}$ at T=50~mK for a small range of BG voltages. This data was obtained after the BG was first swept up to 22.5~V and then reduced to 7.245~V. As described above, at $V_{L\&R}=0$~V the sample is now less conductive, but sweeping $V_{L\&R}$ to positive values increases $I_{sd}$. It is also evident that there is a finer structure that emerges at these low temperatures, which is clearly absent at T=4.2~K (Figure~\ref{fig4}a). These modulations in conductance are highly reproducible and show a continuous evolution with $V_{BG}$. We attribute this structure to mesoscopic effects that arise from disorder in the conducting channel. We believe that the integration of such split-gate devices with higher mobility oxide interfaces with significantly longer mean free paths~\cite{Chen_Al2O3_NatComm, Huijben_Adv_Fun_Mat} should enable studies of mesoscopic transport in the 1D limit.

Thus far we have concentrated on the flow of normal electrons through devices defined via local top gates. Of course, one of the remarkable properties of the LAO/STO interface is that it can also host 2D-superconductivity~\cite{Reyren_Science}. Furthermore, bulk studies have shown that reducing the carrier density at the interface with a global back gate results in a quantum phase transition between superconducting and insulating ground states~\cite{Caviglia_Nature}. The ability to \emph{locally} alter the ground state of the interface at the nanoscale provides the opportunity to create gate-tunable superconducting circuit elements (e.g., Josephson junctions), which may enhance our understanding of the microscopic nature of superconductivity at the interface. To study the effects of a local top gate on the superconductivity we cooled down Dev2 (see Figure~\ref{fig1}c for SEM image) to T=50~mK, and recorded current-voltage (I-V) characteristics as a function of $V_{gate}$ in a four probe configuration (with BG fixed at zero). In Figure~\ref{fig5}a we plot a few representative I-V traces at different values of $V_{gate}$ (inset shows a to-scale schematic of the measurement configuration). These curves show a clear transition from a superconducting state ($V_{gate}=0$~V, black trace) to an insulating state as $V_{gate}$ is made more negative. This insulating state is evident from a gap-like structure that emerges for $V_{gate}<-0.9$~V, and grows in size as the gate voltage is further reduced. Though previous studies with large top gates have shown some gate dependent modulation of the critical current~\cite{Eerkes_APL}, we believe that a local superconducting-insulating transition has thus far not been observed at the LAO/STO interface.

Figure~\ref{fig5}b shows a 2D plot of differential resistance $dV/dI$ (obtained via numerical differentiation of the gate-dependent I-V traces) as a function of $V_{gate}$ and $I$. Two important features to note in this plot are (i) strong peaks that occur at relatively high currents (10 - 15~nA) and (ii) a weaker peak at much lower currents ($<3$~nA), which closes as $V_{gate}$ is made more negative and finally disappears around the superconducting-insulating transition. The stronger set of peaks reflect the current driven superconducting to normal transition in the areas outside the top gated region (we comment further on them toward the end). To understand the origin of the strongly gate-dependent structure at lower currents, we explore the possibility that the region under the top gate acts as a weak link between the superconducting reservoirs on either side. Figure~\ref{fig5}c shows the $dV/dI$ map in a smaller range (indicated by dashed rectangle in Figure~\ref{fig5}b) with two representative line traces at $V_{gate}=0$~V and $V_{gate}=-0.6$~V. We start by associating a critical current ($I_c$) with the first local maximum in $dV/dI$. Figure~\ref{fig5}d (lower panel) shows that $I_c$ is roughly constant from $V_{gate}=0$~V to $V_{gate}=-0.5$~V, below which it begins to drop more rapidly and finally disappears at $V_{gate}=-0.92$~V (red line is a guide to the eye). In the upper panel we plot the product of $I_c$ and $R_n$ (where $R_n$ is the resistance just above $I_c$), which remains roughly constant over the entire gate voltage range, a characteristic signature of a Josephson junction. For $T<<T_c$ ($T_c$ is the critical temperature of the superconducting reservoirs), we expect $eI_cR_n\approx\alpha\Delta$, where $\Delta$ is the superconducting gap in the reservoirs, $e$ is the electronic charge, and $\alpha$ depends on the microscopic details of the weak link and can take values from $\sim 1.5-3$~\cite{Likharev_RMP}. The weak link can in general be a tunnel barrier, a superconductor or a normal metal. Interestingly, for the LAO/STO system all three scenarios are possible, and the current experiments cannot precisely determine the actual microscopic nature of the weak link. However, using the fact that we experimentally determine $eI_cR_n\sim 30\mu$eV, we estimate that $\Delta\sim 10-20\mu$eV. Though these values are slightly smaller (roughly by a factor of 3) than recent measurements of $\Delta$ via tunneling spectroscopy~\cite{Richter_Nature_Gap}, they are consistent with the the lower $T_c\sim 100$~mK of our samples (see SI). We note that similar observations of a gate-tunable weak link at the surface of STO have also recently been reported~\cite{Goldhaber_JJ}.

Finally, we remark on the evolution of the peaks in $dV/dI$ at higher values of $I$ and associated with superconducting-normal switching of the area outside the top gated region. We associate the multiple switches with inhomogeneities in the long (150~$\mu$m) and narrow (5~$\mu$m) conducting channel across which the measurements were performed. Most of the peaks run parallel to each other, showing hardly any gate-dependent shifts till about $V_{gate}=-0.9$~V, which is close to where the superconducting-insulating transition occurs. Interestingly, below $V_{gate}=-0.9$~V these peaks start coming together more rapidly. Though we do not have a complete understanding of this behavior at the moment, we present two possible explanations. The first possibility is that the significantly higher resistance of the insulating state gives rise to Joule heating which increases the local electronic temperature, thereby reducing the critical current of the neighboring regions. The second possibility relates to the earlier discussion (in the context of split gates), whereby electric field lines from the top gate could potentially spread out significantly once the density below the gate is reduced. In this scenario, the evolution of the peaks could be explained by the fact that the electrostatic region of influence of the top gate extends significantly beyond its geometric dimensions, thereby reducing the critical current in the neighboring regions. In contrast to our current device geometry, we believe that devices with significantly wider superconducting banks would be more suitable for the study of the Josephson effect. In such devices the un-gated regions would have a much larger critical current than the weak link, and the strong influence of local inhomogeneities could possibly be suppressed, thereby allowing for a clearer interpretation of the results.

In conclusion, we have demonstrated that top gating can be used to create electrostatically confined nanostructures at the LAO/STO interface. These gates show excellent performance and stability from room temperature down to 50~mK. Not only do they allow us to control the electrostatic landscape through which normal electrons flow, they provide a promising route to locally control the electronic ground state of the interface. The inherent flexibility in the design of such top-gated structures opens up a new and versatile platform for creating a variety of gate-tunable nanostructures at oxide interfaces.

We thank A.~Akhmerov, A.~Geresdi, P.~Gallagher and D.~Goldhaber-Gordon for helpful discussions and insights, P.~D.~Eerkes and C.~Richter for inputs regarding device fabrication, M.~van~Oossanen and T.~Kool for technical assistance, and V.~E.~Calado for help with SEM imaging. This work was supported by the Netherlands Organisation for Scientific Research (NWO/OCW) as part of the Frontiers of Nanoscience program, and by the Dutch Foundation for Fundamental Research on Matter (FOM).

\bibliography{conf_ox_refs}

\begin{thebibliography}{37}
\expandafter\ifx\csname natexlab\endcsname\relax\def\natexlab#1{#1}\fi
\expandafter\ifx\csname bibnamefont\endcsname\relax
  \def\bibnamefont#1{#1}\fi
\expandafter\ifx\csname bibfnamefont\endcsname\relax
  \def\bibfnamefont#1{#1}\fi
\expandafter\ifx\csname citenamefont\endcsname\relax
  \def\citenamefont#1{#1}\fi
\expandafter\ifx\csname url\endcsname\relax
  \def\url#1{\texttt{#1}}\fi
\expandafter\ifx\csname urlprefix\endcsname\relax\def\urlprefix{URL }\fi
\providecommand{\bibinfo}[2]{#2}
\providecommand{\eprint}[2][]{\url{#2}}

\bibitem[{\citenamefont{Dagotto}(2005)}]{Dagotto_Science_Rev}
\bibinfo{author}{\bibfnamefont{E.}~\bibnamefont{Dagotto}},
  \bibinfo{journal}{Science} \textbf{\bibinfo{volume}{309}},
  \bibinfo{pages}{257} (\bibinfo{year}{2005}).

\bibitem[{\citenamefont{Zubko et~al.}(2011)\citenamefont{Zubko, Gariglio,
  Gabay, Ghosez, and Triscone}}]{Zubko_Triscone_ARCMP}
\bibinfo{author}{\bibfnamefont{P.}~\bibnamefont{Zubko}},
  \bibinfo{author}{\bibfnamefont{S.}~\bibnamefont{Gariglio}},
  \bibinfo{author}{\bibfnamefont{M.}~\bibnamefont{Gabay}},
  \bibinfo{author}{\bibfnamefont{P.}~\bibnamefont{Ghosez}}, \bibnamefont{and}
  \bibinfo{author}{\bibfnamefont{J.-M.} \bibnamefont{Triscone}},
  \bibinfo{journal}{Annu. Rev. Condens. Matter Phys.}
  \textbf{\bibinfo{volume}{2}}, \bibinfo{pages}{141} (\bibinfo{year}{2011}).

\bibitem[{\citenamefont{Hwang et~al.}(2012)\citenamefont{Hwang, Iwasa,
  Kawasaki, Keimer, Nagaosa, and Tokura}}]{Hwang_NatMat_Rev}
\bibinfo{author}{\bibfnamefont{H.~Y.} \bibnamefont{Hwang}},
  \bibinfo{author}{\bibfnamefont{Y.}~\bibnamefont{Iwasa}},
  \bibinfo{author}{\bibfnamefont{M.}~\bibnamefont{Kawasaki}},
  \bibinfo{author}{\bibfnamefont{B.}~\bibnamefont{Keimer}},
  \bibinfo{author}{\bibfnamefont{N.}~\bibnamefont{Nagaosa}}, \bibnamefont{and}
  \bibinfo{author}{\bibfnamefont{Y.}~\bibnamefont{Tokura}},
  \bibinfo{journal}{Nat. Mater.} \textbf{\bibinfo{volume}{11}},
  \bibinfo{pages}{103} (\bibinfo{year}{2012}).

\bibitem[{\citenamefont{Ohtomo and Hwang}(2004)}]{Ohtomo_Nature}
\bibinfo{author}{\bibfnamefont{A.}~\bibnamefont{Ohtomo}} \bibnamefont{and}
  \bibinfo{author}{\bibfnamefont{H.~Y.} \bibnamefont{Hwang}},
  \bibinfo{journal}{Nature} \textbf{\bibinfo{volume}{427}},
  \bibinfo{pages}{423} (\bibinfo{year}{2004}).

\bibitem[{\citenamefont{Chen et~al.}(2013)\citenamefont{Chen, Bovet, Trier,
  Christensen, Qu, Andersen, Kasama, Zhang, Giraud, Dufouleur
  et~al.}}]{Chen_Al2O3_NatComm}
\bibinfo{author}{\bibfnamefont{Y.~Z.} \bibnamefont{Chen}},
  \bibinfo{author}{\bibfnamefont{N.}~\bibnamefont{Bovet}},
  \bibinfo{author}{\bibfnamefont{F.}~\bibnamefont{Trier}},
  \bibinfo{author}{\bibfnamefont{D.~V.} \bibnamefont{Christensen}},
  \bibinfo{author}{\bibfnamefont{F.~M.} \bibnamefont{Qu}},
  \bibinfo{author}{\bibfnamefont{N.~H.} \bibnamefont{Andersen}},
  \bibinfo{author}{\bibfnamefont{T.}~\bibnamefont{Kasama}},
  \bibinfo{author}{\bibfnamefont{W.}~\bibnamefont{Zhang}},
  \bibinfo{author}{\bibfnamefont{R.}~\bibnamefont{Giraud}},
  \bibinfo{author}{\bibfnamefont{J.}~\bibnamefont{Dufouleur}},
  \bibnamefont{et~al.}, \bibinfo{journal}{Nat. Comm.}
  \textbf{\bibinfo{volume}{4}}, \bibinfo{pages}{1371} (\bibinfo{year}{2013}).

\bibitem[{\citenamefont{Caviglia
  et~al.}(2010{\natexlab{a}})\citenamefont{Caviglia, Gariglio, Cancellieri,
  Sac\'ep\'e, F\^ete, Reyren, Gabay, Morpurgo, and
  Triscone}}]{Caviglia_SdH_PRL}
\bibinfo{author}{\bibfnamefont{A.~D.} \bibnamefont{Caviglia}},
  \bibinfo{author}{\bibfnamefont{S.}~\bibnamefont{Gariglio}},
  \bibinfo{author}{\bibfnamefont{C.}~\bibnamefont{Cancellieri}},
  \bibinfo{author}{\bibfnamefont{B.}~\bibnamefont{Sac\'ep\'e}},
  \bibinfo{author}{\bibfnamefont{A.}~\bibnamefont{F\^ete}},
  \bibinfo{author}{\bibfnamefont{N.}~\bibnamefont{Reyren}},
  \bibinfo{author}{\bibfnamefont{M.}~\bibnamefont{Gabay}},
  \bibinfo{author}{\bibfnamefont{A.~F.} \bibnamefont{Morpurgo}},
  \bibnamefont{and} \bibinfo{author}{\bibfnamefont{J.-M.}
  \bibnamefont{Triscone}}, \bibinfo{journal}{Phys. Rev. Lett.}
  \textbf{\bibinfo{volume}{105}}, \bibinfo{pages}{236802}
  (\bibinfo{year}{2010}{\natexlab{a}}).

\bibitem[{\citenamefont{Huijben et~al.}(2013)\citenamefont{Huijben, Koster,
  Kruize, Wenderich, Verbeeck, Bals, Slooten, Shi, Molegraaf, Kleibeuker
  et~al.}}]{Huijben_Adv_Fun_Mat}
\bibinfo{author}{\bibfnamefont{M.}~\bibnamefont{Huijben}},
  \bibinfo{author}{\bibfnamefont{G.}~\bibnamefont{Koster}},
  \bibinfo{author}{\bibfnamefont{M.~K.} \bibnamefont{Kruize}},
  \bibinfo{author}{\bibfnamefont{S.}~\bibnamefont{Wenderich}},
  \bibinfo{author}{\bibfnamefont{J.}~\bibnamefont{Verbeeck}},
  \bibinfo{author}{\bibfnamefont{S.}~\bibnamefont{Bals}},
  \bibinfo{author}{\bibfnamefont{E.}~\bibnamefont{Slooten}},
  \bibinfo{author}{\bibfnamefont{B.}~\bibnamefont{Shi}},
  \bibinfo{author}{\bibfnamefont{H.~J.~A.} \bibnamefont{Molegraaf}},
  \bibinfo{author}{\bibfnamefont{J.~E.} \bibnamefont{Kleibeuker}},
  \bibnamefont{et~al.}, \bibinfo{journal}{Adv. Funct. Mater.}
  \textbf{\bibinfo{volume}{23}}, \bibinfo{pages}{5240} (\bibinfo{year}{2013}).

\bibitem[{\citenamefont{Brinkman et~al.}(2007)\citenamefont{Brinkman, Huijben,
  Van~Zalk, Huijben, Zeitler, Maan, Van~der Wiel, Rijnders, Blank, and
  Hilgenkamp}}]{Brinkman_Nat_Mat}
\bibinfo{author}{\bibfnamefont{A.}~\bibnamefont{Brinkman}},
  \bibinfo{author}{\bibfnamefont{M.}~\bibnamefont{Huijben}},
  \bibinfo{author}{\bibfnamefont{M.}~\bibnamefont{Van~Zalk}},
  \bibinfo{author}{\bibfnamefont{J.}~\bibnamefont{Huijben}},
  \bibinfo{author}{\bibfnamefont{U.}~\bibnamefont{Zeitler}},
  \bibinfo{author}{\bibfnamefont{J.~C.} \bibnamefont{Maan}},
  \bibinfo{author}{\bibfnamefont{W.~G.} \bibnamefont{Van~der Wiel}},
  \bibinfo{author}{\bibfnamefont{G.}~\bibnamefont{Rijnders}},
  \bibinfo{author}{\bibfnamefont{D.~H.~A.} \bibnamefont{Blank}},
  \bibnamefont{and}
  \bibinfo{author}{\bibfnamefont{H.}~\bibnamefont{Hilgenkamp}},
  \bibinfo{journal}{Nat. Mater.} \textbf{\bibinfo{volume}{6}},
  \bibinfo{pages}{493} (\bibinfo{year}{2007}).

\bibitem[{\citenamefont{Reyren et~al.}(2007)\citenamefont{Reyren, Thiel,
  Caviglia, Kourkoutis, Hammerl, Richter, Schneider, Kopp, Rüetschi, Jaccard
  et~al.}}]{Reyren_Science}
\bibinfo{author}{\bibfnamefont{N.}~\bibnamefont{Reyren}},
  \bibinfo{author}{\bibfnamefont{S.}~\bibnamefont{Thiel}},
  \bibinfo{author}{\bibfnamefont{A.~D.} \bibnamefont{Caviglia}},
  \bibinfo{author}{\bibfnamefont{L.~F.} \bibnamefont{Kourkoutis}},
  \bibinfo{author}{\bibfnamefont{G.}~\bibnamefont{Hammerl}},
  \bibinfo{author}{\bibfnamefont{C.}~\bibnamefont{Richter}},
  \bibinfo{author}{\bibfnamefont{C.~W.} \bibnamefont{Schneider}},
  \bibinfo{author}{\bibfnamefont{T.}~\bibnamefont{Kopp}},
  \bibinfo{author}{\bibfnamefont{A.-S.} \bibnamefont{Rüetschi}},
  \bibinfo{author}{\bibfnamefont{D.}~\bibnamefont{Jaccard}},
  \bibnamefont{et~al.}, \bibinfo{journal}{Science}
  \textbf{\bibinfo{volume}{317}}, \bibinfo{pages}{1196} (\bibinfo{year}{2007}).

\bibitem[{\citenamefont{Thiel et~al.}(2006)\citenamefont{Thiel, Hammerl,
  Schmehl, Schneider, and Mannhart}}]{Thiel_Science}
\bibinfo{author}{\bibfnamefont{S.}~\bibnamefont{Thiel}},
  \bibinfo{author}{\bibfnamefont{G.}~\bibnamefont{Hammerl}},
  \bibinfo{author}{\bibfnamefont{A.}~\bibnamefont{Schmehl}},
  \bibinfo{author}{\bibfnamefont{C.~W.} \bibnamefont{Schneider}},
  \bibnamefont{and} \bibinfo{author}{\bibfnamefont{J.}~\bibnamefont{Mannhart}},
  \bibinfo{journal}{Science} \textbf{\bibinfo{volume}{313}},
  \bibinfo{pages}{1942} (\bibinfo{year}{2006}).

\bibitem[{\citenamefont{Caviglia
  et~al.}(2010{\natexlab{b}})\citenamefont{Caviglia, Gabay, Gariglio, Reyren,
  Cancellieri, and Triscone}}]{Caviglia_SO_PRL}
\bibinfo{author}{\bibfnamefont{A.~D.} \bibnamefont{Caviglia}},
  \bibinfo{author}{\bibfnamefont{M.}~\bibnamefont{Gabay}},
  \bibinfo{author}{\bibfnamefont{S.}~\bibnamefont{Gariglio}},
  \bibinfo{author}{\bibfnamefont{N.}~\bibnamefont{Reyren}},
  \bibinfo{author}{\bibfnamefont{C.}~\bibnamefont{Cancellieri}},
  \bibnamefont{and} \bibinfo{author}{\bibfnamefont{J.-M.}
  \bibnamefont{Triscone}}, \bibinfo{journal}{Phys. Rev. Lett.}
  \textbf{\bibinfo{volume}{104}}, \bibinfo{pages}{126803}
  (\bibinfo{year}{2010}{\natexlab{b}}).

\bibitem[{\citenamefont{Ben~Shalom et~al.}(2010)\citenamefont{Ben~Shalom,
  Sachs, Rakhmilevitch, Palevski, and Dagan}}]{Shalom_SO_PRL}
\bibinfo{author}{\bibfnamefont{M.}~\bibnamefont{Ben~Shalom}},
  \bibinfo{author}{\bibfnamefont{M.}~\bibnamefont{Sachs}},
  \bibinfo{author}{\bibfnamefont{D.}~\bibnamefont{Rakhmilevitch}},
  \bibinfo{author}{\bibfnamefont{A.}~\bibnamefont{Palevski}}, \bibnamefont{and}
  \bibinfo{author}{\bibfnamefont{Y.}~\bibnamefont{Dagan}},
  \bibinfo{journal}{Phys. Rev. Lett.} \textbf{\bibinfo{volume}{104}},
  \bibinfo{pages}{126802} (\bibinfo{year}{2010}).

\bibitem[{\citenamefont{Caviglia et~al.}(2008)\citenamefont{Caviglia, Gariglio,
  Reyren, Jaccard, Schneider, Gabay, Thiel, Hammerl, Mannhart, and
  Triscone}}]{Caviglia_Nature}
\bibinfo{author}{\bibfnamefont{A.~D.} \bibnamefont{Caviglia}},
  \bibinfo{author}{\bibfnamefont{S.}~\bibnamefont{Gariglio}},
  \bibinfo{author}{\bibfnamefont{N.}~\bibnamefont{Reyren}},
  \bibinfo{author}{\bibfnamefont{D.}~\bibnamefont{Jaccard}},
  \bibinfo{author}{\bibfnamefont{T.}~\bibnamefont{Schneider}},
  \bibinfo{author}{\bibfnamefont{M.}~\bibnamefont{Gabay}},
  \bibinfo{author}{\bibfnamefont{S.}~\bibnamefont{Thiel}},
  \bibinfo{author}{\bibfnamefont{G.}~\bibnamefont{Hammerl}},
  \bibinfo{author}{\bibfnamefont{J.}~\bibnamefont{Mannhart}}, \bibnamefont{and}
  \bibinfo{author}{\bibfnamefont{J.~M.} \bibnamefont{Triscone}},
  \bibinfo{journal}{Nature} \textbf{\bibinfo{volume}{456}},
  \bibinfo{pages}{624} (\bibinfo{year}{2008}).

\bibitem[{\citenamefont{Bert et~al.}(2011)\citenamefont{Bert, Kalisky, Bell,
  Kim, Hikita, Hwang, and Moler}}]{Bert_Moler_Nat_Phys}
\bibinfo{author}{\bibfnamefont{J.~A.} \bibnamefont{Bert}},
  \bibinfo{author}{\bibfnamefont{B.}~\bibnamefont{Kalisky}},
  \bibinfo{author}{\bibfnamefont{C.}~\bibnamefont{Bell}},
  \bibinfo{author}{\bibfnamefont{M.}~\bibnamefont{Kim}},
  \bibinfo{author}{\bibfnamefont{Y.}~\bibnamefont{Hikita}},
  \bibinfo{author}{\bibfnamefont{H.~Y.} \bibnamefont{Hwang}}, \bibnamefont{and}
  \bibinfo{author}{\bibfnamefont{K.~A.} \bibnamefont{Moler}},
  \bibinfo{journal}{Nat. Phys.} \textbf{\bibinfo{volume}{7}},
  \bibinfo{pages}{767} (\bibinfo{year}{2011}).

\bibitem[{\citenamefont{Kalisky et~al.}(2013)\citenamefont{Kalisky, Spanton,
  Noad, Kirtley, Nowack, Bell, Sato, Hosoda, Xie, Hikita
  et~al.}}]{Kalisky_Moler_Nat_Mat}
\bibinfo{author}{\bibfnamefont{B.}~\bibnamefont{Kalisky}},
  \bibinfo{author}{\bibfnamefont{E.~M.} \bibnamefont{Spanton}},
  \bibinfo{author}{\bibfnamefont{H.}~\bibnamefont{Noad}},
  \bibinfo{author}{\bibfnamefont{J.~R.} \bibnamefont{Kirtley}},
  \bibinfo{author}{\bibfnamefont{K.~C.} \bibnamefont{Nowack}},
  \bibinfo{author}{\bibfnamefont{C.}~\bibnamefont{Bell}},
  \bibinfo{author}{\bibfnamefont{H.~K.} \bibnamefont{Sato}},
  \bibinfo{author}{\bibfnamefont{M.}~\bibnamefont{Hosoda}},
  \bibinfo{author}{\bibfnamefont{Y.}~\bibnamefont{Xie}},
  \bibinfo{author}{\bibfnamefont{Y.}~\bibnamefont{Hikita}},
  \bibnamefont{et~al.}, \bibinfo{journal}{Nat. Mater.}
  \textbf{\bibinfo{volume}{12}}, \bibinfo{pages}{1091} (\bibinfo{year}{2013}).

\bibitem[{\citenamefont{Honig et~al.}(2013)\citenamefont{Honig, Sulpizio,
  Drori, Joshua, Zeldov, and Ilani}}]{Honig_Ilani_Nat_Mat}
\bibinfo{author}{\bibfnamefont{M.}~\bibnamefont{Honig}},
  \bibinfo{author}{\bibfnamefont{J.~A.} \bibnamefont{Sulpizio}},
  \bibinfo{author}{\bibfnamefont{J.}~\bibnamefont{Drori}},
  \bibinfo{author}{\bibfnamefont{A.}~\bibnamefont{Joshua}},
  \bibinfo{author}{\bibfnamefont{E.}~\bibnamefont{Zeldov}}, \bibnamefont{and}
  \bibinfo{author}{\bibfnamefont{S.}~\bibnamefont{Ilani}},
  \bibinfo{journal}{Nat. Mater.} \textbf{\bibinfo{volume}{12}},
  \bibinfo{pages}{1112} (\bibinfo{year}{2013}).

\bibitem[{\citenamefont{Fidkowski et~al.}(2013)\citenamefont{Fidkowski, Jiang,
  Lutchyn, and Nayak}}]{Fidkowski_Majo_PRB}
\bibinfo{author}{\bibfnamefont{L.}~\bibnamefont{Fidkowski}},
  \bibinfo{author}{\bibfnamefont{H.-C.} \bibnamefont{Jiang}},
  \bibinfo{author}{\bibfnamefont{R.~M.} \bibnamefont{Lutchyn}},
  \bibnamefont{and} \bibinfo{author}{\bibfnamefont{C.}~\bibnamefont{Nayak}},
  \bibinfo{journal}{Phys. Rev. B} \textbf{\bibinfo{volume}{87}},
  \bibinfo{pages}{014436} (\bibinfo{year}{2013}).

\bibitem[{\citenamefont{Mannhart and Schlom}(2010)}]{Mannhart_Science_Rev}
\bibinfo{author}{\bibfnamefont{J.}~\bibnamefont{Mannhart}} \bibnamefont{and}
  \bibinfo{author}{\bibfnamefont{D.~G.} \bibnamefont{Schlom}},
  \bibinfo{journal}{Science} \textbf{\bibinfo{volume}{327}},
  \bibinfo{pages}{1607} (\bibinfo{year}{2010}).

\bibitem[{\citenamefont{Stornaiuolo et~al.}(2012)\citenamefont{Stornaiuolo,
  Gariglio, Couto, F\^{e}te, Caviglia, Seyfarth, Jaccard, Morpurgo, and
  Triscone}}]{Stornaiuolo_confinement_APL}
\bibinfo{author}{\bibfnamefont{D.}~\bibnamefont{Stornaiuolo}},
  \bibinfo{author}{\bibfnamefont{S.}~\bibnamefont{Gariglio}},
  \bibinfo{author}{\bibfnamefont{N.~J.~G.} \bibnamefont{Couto}},
  \bibinfo{author}{\bibfnamefont{A.}~\bibnamefont{F\^{e}te}},
  \bibinfo{author}{\bibfnamefont{A.~D.} \bibnamefont{Caviglia}},
  \bibinfo{author}{\bibfnamefont{G.}~\bibnamefont{Seyfarth}},
  \bibinfo{author}{\bibfnamefont{D.}~\bibnamefont{Jaccard}},
  \bibinfo{author}{\bibfnamefont{A.~F.} \bibnamefont{Morpurgo}},
  \bibnamefont{and} \bibinfo{author}{\bibfnamefont{J.-M.}
  \bibnamefont{Triscone}}, \bibinfo{journal}{Appl. Phys. Lett.}
  \textbf{\bibinfo{volume}{101}}, \bibinfo{eid}{222601} (\bibinfo{year}{2012}).

\bibitem[{\citenamefont{Paolo~Aurino et~al.}(2013)\citenamefont{Paolo~Aurino,
  Kalabukhov, Tuzla, Olsson, Claeson, and Winkler}}]{Ionbeam_APL}
\bibinfo{author}{\bibfnamefont{P.}~\bibnamefont{Paolo~Aurino}},
  \bibinfo{author}{\bibfnamefont{A.}~\bibnamefont{Kalabukhov}},
  \bibinfo{author}{\bibfnamefont{N.}~\bibnamefont{Tuzla}},
  \bibinfo{author}{\bibfnamefont{E.}~\bibnamefont{Olsson}},
  \bibinfo{author}{\bibfnamefont{T.}~\bibnamefont{Claeson}}, \bibnamefont{and}
  \bibinfo{author}{\bibfnamefont{D.}~\bibnamefont{Winkler}},
  \bibinfo{journal}{Applied Physics Letters} \textbf{\bibinfo{volume}{102}},
  \bibinfo{eid}{201610} (\bibinfo{year}{2013}).

\bibitem[{\citenamefont{Cen et~al.}(2008)\citenamefont{Cen, Thiel, Hammerl,
  Schneider, Andersen, Hellberg, Mannhart, and Levy}}]{Cen_NatMat}
\bibinfo{author}{\bibfnamefont{C.}~\bibnamefont{Cen}},
  \bibinfo{author}{\bibfnamefont{S.}~\bibnamefont{Thiel}},
  \bibinfo{author}{\bibfnamefont{G.}~\bibnamefont{Hammerl}},
  \bibinfo{author}{\bibfnamefont{C.~W.} \bibnamefont{Schneider}},
  \bibinfo{author}{\bibfnamefont{K.~E.} \bibnamefont{Andersen}},
  \bibinfo{author}{\bibfnamefont{C.~S.} \bibnamefont{Hellberg}},
  \bibinfo{author}{\bibfnamefont{J.}~\bibnamefont{Mannhart}}, \bibnamefont{and}
  \bibinfo{author}{\bibfnamefont{J.}~\bibnamefont{Levy}},
  \bibinfo{journal}{Nat. Mater.} \textbf{\bibinfo{volume}{7}},
  \bibinfo{pages}{298} (\bibinfo{year}{2008}).

\bibitem[{\citenamefont{Cen et~al.}(2009)\citenamefont{Cen, Thiel, Mannhart,
  and Levy}}]{Cen_Science}
\bibinfo{author}{\bibfnamefont{C.}~\bibnamefont{Cen}},
  \bibinfo{author}{\bibfnamefont{S.}~\bibnamefont{Thiel}},
  \bibinfo{author}{\bibfnamefont{J.}~\bibnamefont{Mannhart}}, \bibnamefont{and}
  \bibinfo{author}{\bibfnamefont{J.}~\bibnamefont{Levy}},
  \bibinfo{journal}{Science} \textbf{\bibinfo{volume}{323}},
  \bibinfo{pages}{1026} (\bibinfo{year}{2009}).

\bibitem[{\citenamefont{Cheng et~al.}(2011)\citenamefont{Cheng, Siles, Bi, Cen,
  Bogorin, Bark, Folkman, Park, Eom, Medeiros-Ribeiro
  et~al.}}]{Cheng_Levy_SET_NatNano}
\bibinfo{author}{\bibfnamefont{G.}~\bibnamefont{Cheng}},
  \bibinfo{author}{\bibfnamefont{P.~F.} \bibnamefont{Siles}},
  \bibinfo{author}{\bibfnamefont{F.}~\bibnamefont{Bi}},
  \bibinfo{author}{\bibfnamefont{C.}~\bibnamefont{Cen}},
  \bibinfo{author}{\bibfnamefont{D.~F.} \bibnamefont{Bogorin}},
  \bibinfo{author}{\bibfnamefont{C.~W.} \bibnamefont{Bark}},
  \bibinfo{author}{\bibfnamefont{C.~M.} \bibnamefont{Folkman}},
  \bibinfo{author}{\bibfnamefont{J.-W.} \bibnamefont{Park}},
  \bibinfo{author}{\bibfnamefont{C.-B.} \bibnamefont{Eom}},
  \bibinfo{author}{\bibfnamefont{G.}~\bibnamefont{Medeiros-Ribeiro}},
  \bibnamefont{et~al.}, \bibinfo{journal}{Nat. Nano.}
  \textbf{\bibinfo{volume}{6}}, \bibinfo{pages}{343} (\bibinfo{year}{2011}).

\bibitem[{\citenamefont{Hosoda et~al.}(2013)\citenamefont{Hosoda, Hikita,
  Hwang, and Bell}}]{Hosada_APL}
\bibinfo{author}{\bibfnamefont{M.}~\bibnamefont{Hosoda}},
  \bibinfo{author}{\bibfnamefont{Y.}~\bibnamefont{Hikita}},
  \bibinfo{author}{\bibfnamefont{H.~Y.} \bibnamefont{Hwang}}, \bibnamefont{and}
  \bibinfo{author}{\bibfnamefont{C.}~\bibnamefont{Bell}},
  \bibinfo{journal}{Appl. Phys. Lett.} \textbf{\bibinfo{volume}{103}},
  \bibinfo{pages}{103507} (\bibinfo{year}{2013}).

\bibitem[{\citenamefont{Eerkes et~al.}(2013)\citenamefont{Eerkes, van~der Wiel,
  and Hilgenkamp}}]{Eerkes_APL}
\bibinfo{author}{\bibfnamefont{P.~D.} \bibnamefont{Eerkes}},
  \bibinfo{author}{\bibfnamefont{W.~G.} \bibnamefont{van~der Wiel}},
  \bibnamefont{and}
  \bibinfo{author}{\bibfnamefont{H.}~\bibnamefont{Hilgenkamp}},
  \bibinfo{journal}{Appl. Phys. Lett.} \textbf{\bibinfo{volume}{103}},
  \bibinfo{pages}{201603} (\bibinfo{year}{2013}).

\bibitem[{\citenamefont{Jany et~al.}(2014)\citenamefont{Jany, Richter,
  Woltmann, Pfanzelt, Förg, Rommel, Reindl, Waizmann, Weis, Mundy
  et~al.}}]{Jany_AdvMat}
\bibinfo{author}{\bibfnamefont{R.}~\bibnamefont{Jany}},
  \bibinfo{author}{\bibfnamefont{C.}~\bibnamefont{Richter}},
  \bibinfo{author}{\bibfnamefont{C.}~\bibnamefont{Woltmann}},
  \bibinfo{author}{\bibfnamefont{G.}~\bibnamefont{Pfanzelt}},
  \bibinfo{author}{\bibfnamefont{B.}~\bibnamefont{Förg}},
  \bibinfo{author}{\bibfnamefont{M.}~\bibnamefont{Rommel}},
  \bibinfo{author}{\bibfnamefont{T.}~\bibnamefont{Reindl}},
  \bibinfo{author}{\bibfnamefont{U.}~\bibnamefont{Waizmann}},
  \bibinfo{author}{\bibfnamefont{J.}~\bibnamefont{Weis}},
  \bibinfo{author}{\bibfnamefont{J.~A.} \bibnamefont{Mundy}},
  \bibnamefont{et~al.}, \bibinfo{journal}{Adv. Mater. Interf.}
  \textbf{\bibinfo{volume}{1}}, \bibinfo{eid}{1300031} (\bibinfo{year}{2014}).

\bibitem[{\citenamefont{F\"org et~al.}(2012)\citenamefont{F\"org, Richter, and
  Mannhart}}]{Forg_APL}
\bibinfo{author}{\bibfnamefont{B.}~\bibnamefont{F\"org}},
  \bibinfo{author}{\bibfnamefont{C.}~\bibnamefont{Richter}}, \bibnamefont{and}
  \bibinfo{author}{\bibfnamefont{J.}~\bibnamefont{Mannhart}},
  \bibinfo{journal}{Appl. Phys. Lett.} \textbf{\bibinfo{volume}{100}},
  \bibinfo{pages}{053506} (\bibinfo{year}{2012}).

\bibitem[{\citenamefont{Richter et~al.}(2013)\citenamefont{Richter, Boschker,
  Dietsche, Fillis-Tsirakis, Jany, Loder, Kourkoutis, Muller, Kirtley,
  Schneider et~al.}}]{Richter_Nature_Gap}
\bibinfo{author}{\bibfnamefont{C.}~\bibnamefont{Richter}},
  \bibinfo{author}{\bibfnamefont{H.}~\bibnamefont{Boschker}},
  \bibinfo{author}{\bibfnamefont{W.}~\bibnamefont{Dietsche}},
  \bibinfo{author}{\bibfnamefont{E.}~\bibnamefont{Fillis-Tsirakis}},
  \bibinfo{author}{\bibfnamefont{R.}~\bibnamefont{Jany}},
  \bibinfo{author}{\bibfnamefont{F.}~\bibnamefont{Loder}},
  \bibinfo{author}{\bibfnamefont{L.~F.} \bibnamefont{Kourkoutis}},
  \bibinfo{author}{\bibfnamefont{D.~A.} \bibnamefont{Muller}},
  \bibinfo{author}{\bibfnamefont{J.~R.} \bibnamefont{Kirtley}},
  \bibinfo{author}{\bibfnamefont{C.~W.} \bibnamefont{Schneider}},
  \bibnamefont{et~al.}, \bibinfo{journal}{Nature}
  \textbf{\bibinfo{volume}{502}}, \bibinfo{pages}{528} (\bibinfo{year}{2013}).

\bibitem[{\citenamefont{van Wees et~al.}(1988)\citenamefont{van Wees, van
  Houten, Beenakker, Williamson, Kouwenhoven, van~der Marel, and
  Foxon}}]{QPC_vanWees_PRL}
\bibinfo{author}{\bibfnamefont{B.~J.} \bibnamefont{van Wees}},
  \bibinfo{author}{\bibfnamefont{H.}~\bibnamefont{van Houten}},
  \bibinfo{author}{\bibfnamefont{C.~W.~J.} \bibnamefont{Beenakker}},
  \bibinfo{author}{\bibfnamefont{J.~G.} \bibnamefont{Williamson}},
  \bibinfo{author}{\bibfnamefont{L.~P.} \bibnamefont{Kouwenhoven}},
  \bibinfo{author}{\bibfnamefont{D.}~\bibnamefont{van~der Marel}},
  \bibnamefont{and} \bibinfo{author}{\bibfnamefont{C.~T.} \bibnamefont{Foxon}},
  \bibinfo{journal}{Phys. Rev. Lett.} \textbf{\bibinfo{volume}{60}},
  \bibinfo{pages}{848} (\bibinfo{year}{1988}).

\bibitem[{\citenamefont{Wharam et~al.}(1988)\citenamefont{Wharam, Thornton,
  Newbury, Pepper, Ahmed, Frost, Hasko, Peacock, Ritchie, and
  Jones}}]{QPC_Wharam}
\bibinfo{author}{\bibfnamefont{D.~A.} \bibnamefont{Wharam}},
  \bibinfo{author}{\bibfnamefont{T.~J.} \bibnamefont{Thornton}},
  \bibinfo{author}{\bibfnamefont{R.}~\bibnamefont{Newbury}},
  \bibinfo{author}{\bibfnamefont{M.}~\bibnamefont{Pepper}},
  \bibinfo{author}{\bibfnamefont{H.}~\bibnamefont{Ahmed}},
  \bibinfo{author}{\bibfnamefont{J.~E.~F.} \bibnamefont{Frost}},
  \bibinfo{author}{\bibfnamefont{D.~G.} \bibnamefont{Hasko}},
  \bibinfo{author}{\bibfnamefont{D.~C.} \bibnamefont{Peacock}},
  \bibinfo{author}{\bibfnamefont{D.~A.} \bibnamefont{Ritchie}},
  \bibnamefont{and} \bibinfo{author}{\bibfnamefont{G.~A.~C.}
  \bibnamefont{Jones}}, \bibinfo{journal}{J. Phys. C}
  \textbf{\bibinfo{volume}{21}}, \bibinfo{pages}{L209} (\bibinfo{year}{1988}).

\bibitem[{\citenamefont{Beenakker and van Houten}(1991)}]{Beenakker_SSP_Rev}
\bibinfo{author}{\bibfnamefont{C.~W.~J.} \bibnamefont{Beenakker}}
  \bibnamefont{and} \bibinfo{author}{\bibfnamefont{H.}~\bibnamefont{van
  Houten}}, \bibinfo{journal}{Solid State Phys.} \textbf{\bibinfo{volume}{44}},
  \bibinfo{pages}{1} (\bibinfo{year}{1991}).

\bibitem[{\citenamefont{Field et~al.}(1993)\citenamefont{Field, Smith, Pepper,
  Ritchie, Frost, Jones, and Hasko}}]{QPC_ChSensor_Field}
\bibinfo{author}{\bibfnamefont{M.}~\bibnamefont{Field}},
  \bibinfo{author}{\bibfnamefont{C.~G.} \bibnamefont{Smith}},
  \bibinfo{author}{\bibfnamefont{M.}~\bibnamefont{Pepper}},
  \bibinfo{author}{\bibfnamefont{D.~A.} \bibnamefont{Ritchie}},
  \bibinfo{author}{\bibfnamefont{J.~E.~F.} \bibnamefont{Frost}},
  \bibinfo{author}{\bibfnamefont{G.~A.~C.} \bibnamefont{Jones}},
  \bibnamefont{and} \bibinfo{author}{\bibfnamefont{D.~G.} \bibnamefont{Hasko}},
  \bibinfo{journal}{Phys. Rev. Lett.} \textbf{\bibinfo{volume}{70}},
  \bibinfo{pages}{1311} (\bibinfo{year}{1993}).

\bibitem[{\citenamefont{Folk et~al.}(2003)\citenamefont{Folk, Potok, Marcus,
  and Umansky}}]{QPC_SpFilter_Folk}
\bibinfo{author}{\bibfnamefont{J.~A.} \bibnamefont{Folk}},
  \bibinfo{author}{\bibfnamefont{R.~M.} \bibnamefont{Potok}},
  \bibinfo{author}{\bibfnamefont{C.~M.} \bibnamefont{Marcus}},
  \bibnamefont{and} \bibinfo{author}{\bibfnamefont{V.}~\bibnamefont{Umansky}},
  \bibinfo{journal}{Science} \textbf{\bibinfo{volume}{299}},
  \bibinfo{pages}{679} (\bibinfo{year}{2003}).

\bibitem[{\citenamefont{Kouwenhoven et~al.}(1997)\citenamefont{Kouwenhoven,
  Marcus, McEuen, Tarucha, Westervelt, and Wingreen}}]{Kouwenhoven_DotReview}
\bibinfo{author}{\bibfnamefont{L.}~\bibnamefont{Kouwenhoven}},
  \bibinfo{author}{\bibfnamefont{C.}~\bibnamefont{Marcus}},
  \bibinfo{author}{\bibfnamefont{P.}~\bibnamefont{McEuen}},
  \bibinfo{author}{\bibfnamefont{S.}~\bibnamefont{Tarucha}},
  \bibinfo{author}{\bibfnamefont{R.}~\bibnamefont{Westervelt}},
  \bibnamefont{and} \bibinfo{author}{\bibfnamefont{N.}~\bibnamefont{Wingreen}},
  in \emph{\bibinfo{booktitle}{Proceedings of the NATO Advanced Study Institute
  on Mesoscopic Electron Transport}} (\bibinfo{publisher}{Kluwer},
  \bibinfo{address}{Dordrecht}, \bibinfo{year}{1997}), vol.
  \bibinfo{volume}{E345}, pp. \bibinfo{pages}{105--214}.

\bibitem[{\citenamefont{Hemberger et~al.}(1995)\citenamefont{Hemberger,
  Lunkenheimer, Viana, B\"ohmer, and Loidl}}]{Edep_dielectric}
\bibinfo{author}{\bibfnamefont{J.}~\bibnamefont{Hemberger}},
  \bibinfo{author}{\bibfnamefont{P.}~\bibnamefont{Lunkenheimer}},
  \bibinfo{author}{\bibfnamefont{R.}~\bibnamefont{Viana}},
  \bibinfo{author}{\bibfnamefont{R.}~\bibnamefont{B\"ohmer}}, \bibnamefont{and}
  \bibinfo{author}{\bibfnamefont{A.}~\bibnamefont{Loidl}},
  \bibinfo{journal}{Phys. Rev. B} \textbf{\bibinfo{volume}{52}},
  \bibinfo{pages}{13159} (\bibinfo{year}{1995}).

\bibitem[{\citenamefont{Likharev}(1979)}]{Likharev_RMP}
\bibinfo{author}{\bibfnamefont{K.~K.} \bibnamefont{Likharev}},
  \bibinfo{journal}{Rev. Mod. Phys.} \textbf{\bibinfo{volume}{51}},
  \bibinfo{pages}{101} (\bibinfo{year}{1979}).

\bibitem[{\citenamefont{Gallagher et~al.}(2014)\citenamefont{Gallagher, Lee,
  Williams, and Goldhaber-Gordon}}]{Goldhaber_JJ}
\bibinfo{author}{\bibfnamefont{P.}~\bibnamefont{Gallagher}},
  \bibinfo{author}{\bibfnamefont{M.}~\bibnamefont{Lee}},
  \bibinfo{author}{\bibfnamefont{J.~R.} \bibnamefont{Williams}},
  \bibnamefont{and}
  \bibinfo{author}{\bibfnamefont{D.}~\bibnamefont{Goldhaber-Gordon}},
  \bibinfo{journal}{Nat. Phys.} \textbf{\bibinfo{volume}{10}},
  \bibinfo{pages}{748} (\bibinfo{year}{2014}).

\end{thebibliography}

\newpage
\setcounter{figure}{0}
\renewcommand{\thefigure}{S\arabic{figure}}

\onecolumngrid
\section{Supplementary Information}

\begin{figure}[!b]
\includegraphics[width=1\linewidth]{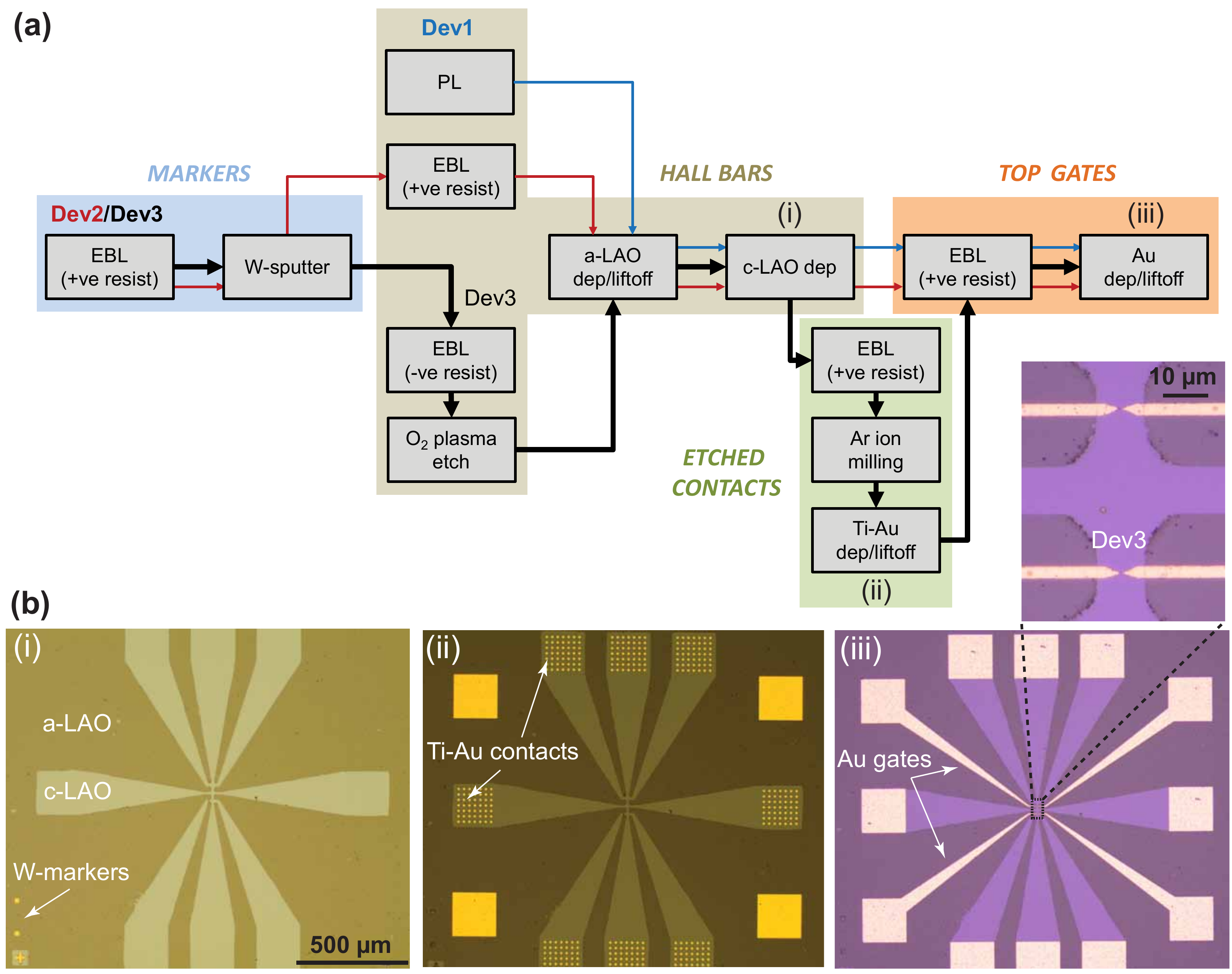}
\caption{(a) Device fabrication flowchart (see text for details). (b) Optical microscope images for Dev3 at various different stages of the fabrication process corresponding to labels (i)-(iii) in (a). The differences in colors between the three images are only a result of different camera settings, and not related to the processes themselves. \label{supfig1}}
\end{figure}

Figure~\ref{supfig1}a shows a complete flowchart of the various steps involved in the fabrication of the devices discussed in the main text (Dev1, Dev2, and Dev3) as well those discussed later in the supplementary information. Bold black arrows show the process flow used to fabricate the split gate (SG) device discussed in detail in the main text (Dev3). Figure~\ref{supfig1}b shows optical images of Dev3 at different stages of the device fabrication corresponding to labels [(i)-(iii)] in Figure~\ref{supfig1}a. The blue/red arrows correspond to Dev1/Dev2 respectively. Dev4, Dev5 and Dev6 (described in Figure~\ref{supfig2}) are fabricated in a similar fashion to Dev3, Dev1, and Dev2 respectively.

\newpage

Below we describe each of the processes in some detail.
\begin{itemize}
\item
\emph{Photolithography (PL)}: PL is performed using a deep-UV Karl Suss MJB3 mask aligner. The sample is coated with S1805 photoresist followed by UV exposure.
\item
\emph{Electron beam lithography (EBL)}: To prepare the sample for EBL, it is typically coated with a positive resist (double layer: PMMA 495K/950K). For Dev3/Dev4 we used a combination of PMMA 495K (100~nm thick) and HSQ (50~nm thick) to define the amorphous LAO (a-LAO) mask. HSQ is a negative resist and therefore allows us to expose only the region that must be protected during the subsequent a-LAO growth. For all EBL steps, the sample is also coated with Aquasave (Mitsubishi Rayon). Aquasave is a water soluble conducting polymer which prevents charging during the lithography process. After lithography, the Aquasave is first dissolved in water, before proceeding with developing. HSQ is developed in TMAH (12.5\% solution in H$_2$O), followed by an H$_2$O and IPA rinse. PMMA is developed in MIBK (33\% solution in IPA) followed by an IPA rinse. Typical doses for the electron beam lithography are 800~$\mu$C/cm$^2$ and 650~$\mu$C/cm$^2$ for PMMA and HSQ respectively.
\item
\emph{Tungsten (W) sputtering}: In order to do well aligned EBL it is necessary to have good quality markers that can survive the high temperatures during the crystalline LAO (c-LAO) growth. We therefore make use of W [see Figure~\ref{supfig1}b-(i)], which is deposited using RF-sputtering at a pressure of 20~$\mu$bar.
\item
\emph{Oxygen plasma etching}: This step is only required if a double layer of PMMA/HSQ was used to define the a-LAO mask. After exposure of the HSQ the oxygen plasma is used remove the PMMA from the unexposed regions. The etching is performed with an RF power of 60~W and oxygen flow rate of 40~sccm for 1~minute.
\item
\emph{Oxides growth}: The details for a-LAO and c-LAO growth have been described in the main text.
\item
\emph{Argon (Ar) ion milling}: Ion milling is used to drill holes that reach the LAO/STO interface. This is followed by in-situ metal deposition, resulting in good ohmic contacts to the interface. The milling is performed in the load lock of the deposition chamber with an Ar pressure of $1\times 10^{-3}$~mbar.
\item
\emph{Metal deposition}: All metals are deposited by electron beam evaporation. Ti/Au (20~nm/50~nm) is deposited after Ar ion milling to make ohmic contacts (in devices where no milling was performed, contacts were made by direct wedge bonding to the interface). For the top gates, only Au (100~nm) was evaporated. The deposition rate was kept low (0.5~A$^\circ$/s) for the first 20~nm after which it was increased to about 2-3~A$^\circ$/s. The base pressure of the deposition chamber is less than 5$\times$10$^{-8}$~mbar and rises to about 1$\times$10$^{-7}$~mbar during the Au deposition.
\item
\emph{Liftoff}: Post-deposition liftoff (a-LAO, W, Ti/Au, and Au) is usually performed in warm acetone (55~$^\circ$C). The process can be sped up by putting the samples in an ultrasonic (US) bath. However, after the top gate deposition (Au) this is avoided in order to prevent the Au from peeling off.
\end{itemize}

\newpage

\begin{figure}[!t]
\includegraphics[width=.6\linewidth]{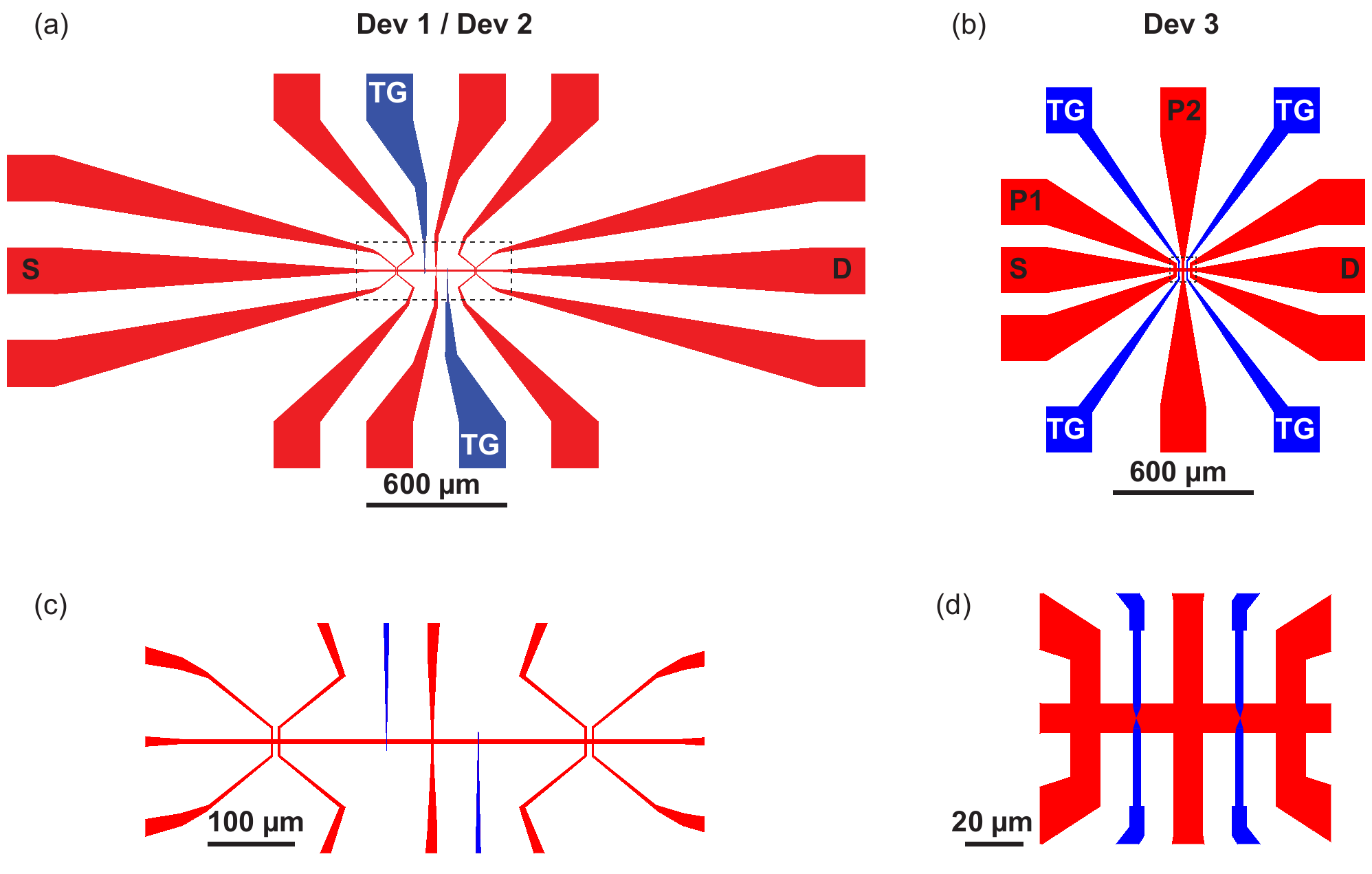}
\caption{(a) Device drawing for Dev1 and Dev2 in main text showing the contacts (S and D) used for two-probe measurements. Red regions correspond to the conducting 2DES and blue regions are the top gates. (b) Similar drawing for the split gate device Dev3 in the main text. P1 and P2 are voltage probes used for four-probe measurements. (c),(d) show close-ups of regions indicated by dashed lines in (a),(b) respectively. The channel widths in the active device region are 5~$\mu$m (Dev1/Dev2) and 10~$\mu$m (Dev3).}
\label{supfig4}
\end{figure}

\begin{figure}[!b]
\includegraphics[width=0.93\linewidth]{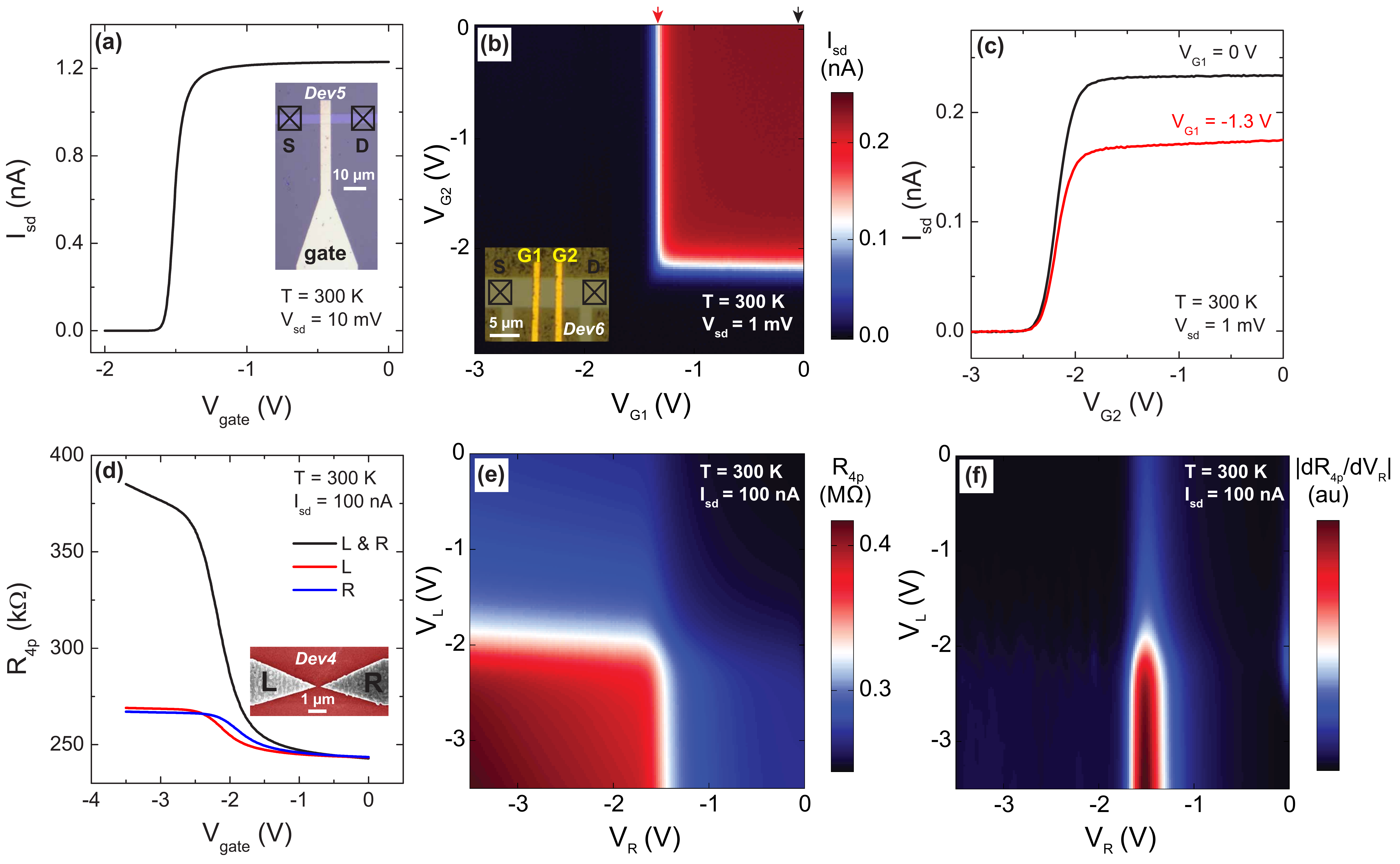}
\caption{(a) Gate characteristics of a 5~$\mu$m wide single gated device (Dev5). (b) 2D map of the current for a device (Dev6) with two gates in series . Inset shows an optical image of the device. (c) Red/black correspond to individual traces taken along the positions marked by red/black arrows in (b). (d) Gate response of a split gate device (Dev4) similar to Dev3, described in the main text. Inset shows SEM image of the device. (e) 2D map of the four-probe resistance ($R_{4p}$) \emph{vs.} the left (L) and right (R) gate voltages for Dev4. (f) Absolute value of the derivative of (e) along the $V_R$ axis.}
\label{supfig2}
\end{figure}

Figure~\ref{supfig2} shows data from three other devices that complement the results described in the main text. Figure~\ref{supfig2}a inset shows a device (Dev5) with a 5~$\mu$m wide gate across the conducting channel. The device characteristics are very similar to those of Dev1 (described in main text), in terms of the on/off ratio and leakage current. We also study transport through a device with two gates in series (Dev6). The 2D current map in Figure~\ref{supfig2}b shows that the two gates act independently of one another, with no cross-talk. Figure~\ref{supfig2}c shows individual traces taken along the positions marked by the red/black arrows in Figure~\ref{supfig2}b. It can clearly be seen that a more negative value of $V_{G1}$ merely reduces the overall conductance of the channel, without affecting the pinch-off voltage for gate G2.

In addition to the split-gate device discussed in the main text (Dev3), we study another device (Dev4) in the same geometry, with a gate separation (tip to tip) of about 250~nm. Figure~\ref{supfig2}d shows the room temperature variation in four-probe resistance ($R_{4p}$) with just the left gate (L), right gate (R), and both together (L\&R). Inset shows a (false color) SEM image of the device. A comparison with Figure~2b (main text) shows qualitatively similar behavior to Dev3, however there is some asymmetry in the action of L and R. This is most likely a result of some local inhomogeneity in the device region. Figure~\ref{supfig2}e shows a 2D map of $R_{4p}$ \emph{vs.} $V_L$, $V_R$. Both the $R_{4p}$ map and the derivative along $V_R$ (Figure~\ref{supfig2}f) look very similar to Dev3 (main text), indicating minimal cross-talk between the gates.

We note that none of the devices studied (in the main text and SI) showed any significant leakage at room temperature. However, we did observe some spread in the pinch-off (threshold) voltage, which varied between -1.5~V and -2.5~V.

\begin{figure}[!b]
\includegraphics[width=0.7\linewidth]{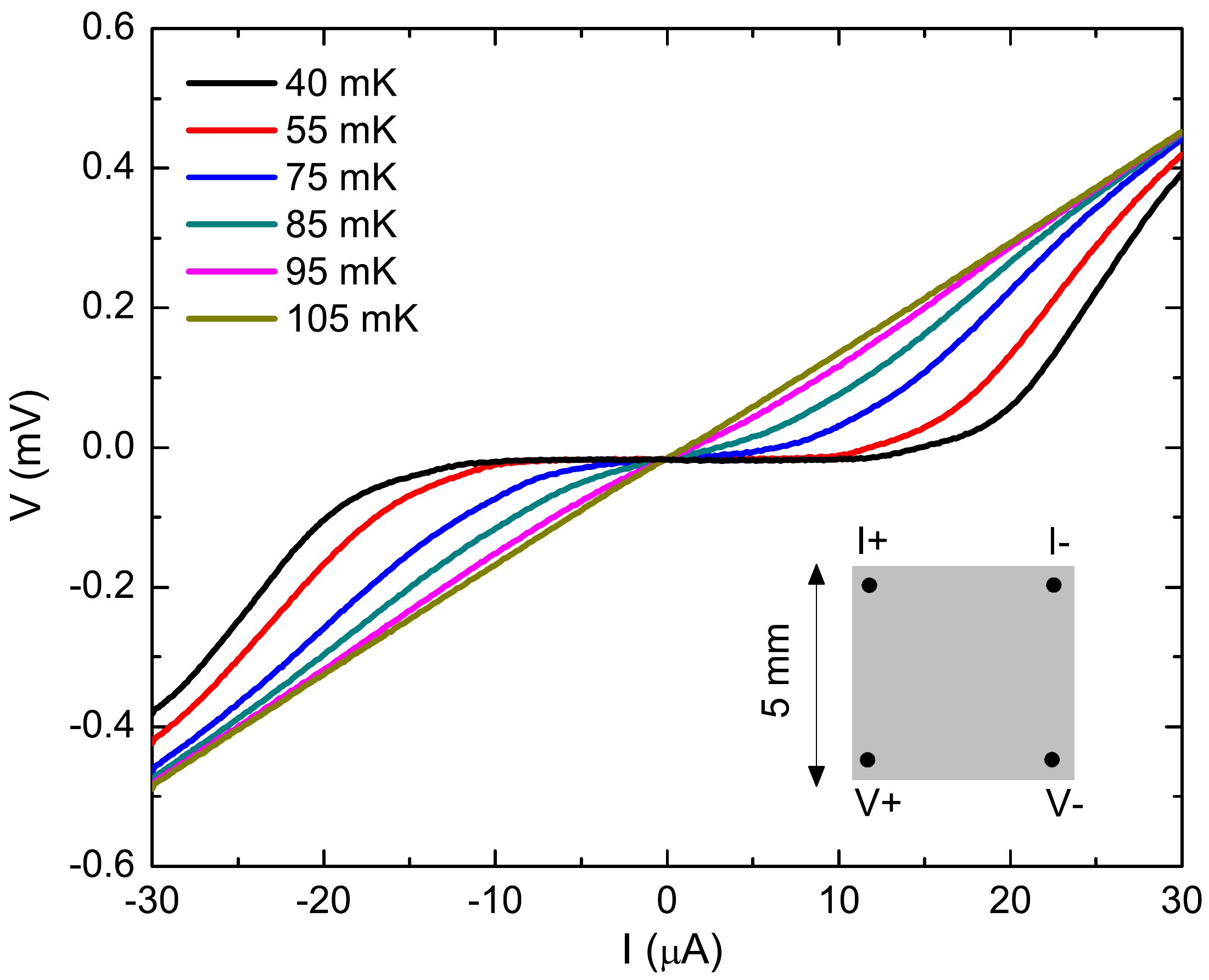}
\caption{I-V characteristics of a bulk sample showing a critical temperature $T_c\sim$100~mK. Measurements were performed in a van der Pauw geometry, as shown in the inset.}
\label{supfig3}
\end{figure}

\end{document}